\documentclass{IEEEtran}
\usepackage[utf8]{inputenc}
\usepackage{amsmath,amssymb,amsfonts}
\usepackage[nospace]{cite}

\usepackage{algorithmic}

\usepackage{graphicx}
\usepackage{textcomp}
\usepackage{xcolor}
\usepackage{pifont}
\usepackage{amsthm}
\usepackage{hyperref}
\usepackage{mathrsfs}
\usepackage{booktabs}
\usepackage{hyperref}
\usepackage{cases}
\usepackage{comment}
\usepackage{empheq} 
\usepackage{caption}
\usepackage{subcaption}
 \usepackage{fancyhdr}

\allowdisplaybreaks

\usepackage{url}  %Required
\usepackage{graphicx}  %Required
\usepackage[ruled,vlined,linesnumbered]{algorithm2e}
\usepackage{setspace}
\usepackage{floatpag} 
\usepackage{float}
\floatpagestyle{empty}

\newtheorem{defn}{Definition}

\usepackage{listings}

\allowdisplaybreaks[4]
\usepackage[english]{babel}
\def\BibTeX{{\rm B\kern-.05em{\sc i\kern-.025em b}\kern-.08em
    T\kern-.1667em\lower.7ex\hbox{E}\kern-.125emX}}
\begin{document}

\title{Privacy-Preserving Blockchain-Based Federated Learning for IoT Devices}

\author{Yang~Zhao,~\IEEEmembership{Student Member,~IEEE,}
        Jun~Zhao,~\IEEEmembership{Member,~IEEE,}
		Linshan~Jiang,~\IEEEmembership{Student Member,~IEEE,}
        Rui~Tan,~\IEEEmembership{Senior Member,~IEEE,}
        Dusit Niyato,~\IEEEmembership{Fellow,~IEEE,}
        Zengxiang~Li,~\IEEEmembership{Member,~IEEE,}
        Lingjuan~Lyu,~\IEEEmembership{Member,~IEEE,}
        and Yingbo~Liu,~\IEEEmembership{Member,~IEEE}
\thanks{
Yang Zhao and Jun Zhao are supported by 1) Nanyang Technological University (NTU) Startup Grant, 2) Alibaba-NTU Singapore Joint Research Institute (JRI), 3) Singapore Ministry of Education Academic Research Fund Tier 1 RG128/18, Tier 1 RG115/19, Tier 1 RT07/19, Tier 1 RT01/19, and Tier 2 MOE2019-T2-1-176, 4) NTU-WASP Joint Project, 5) Singapore National Research Foundation (NRF) under its Strategic Capability Research Centres Funding Initiative: Strategic Centre for Research in Privacy-Preserving Technologies \& Systems (SCRIPTS), 6)  Energy Research Institute @NTU (ERIAN), 7) Singapore NRF National Satellite of Excellence, Design Science and Technology for Secure Critical Infrastructure NSoE DeST-SCI2019-0012, 8) AI Singapore (AISG) 100 Experiments (100E) programme, and 9) NTU Project for Large Vertical Take-Off \& Landing (VTOL) Research Platform. Linshan Jiang and Rui Tan are supported by the MOE AcRF Tier 1 grant 2019-T1-001-044. Dusit Niyato is supported by the National Research Foundation (NRF), Singapore, under Singapore Energy Market Authority (EMA), Energy Resilience, NRF2017EWT-EP003-041, Singapore NRF2015-NRF-ISF001-2277, Singapore NRF National Satellite of Excellence, Design Science and Technology for Secure Critical Infrastructure NSoE DeST-SCI2019-0007, A*STAR-NTU-SUTD Joint Research Grant on Artificial Intelligence for the Future of Manufacturing RGANS1906, Wallenberg AI, Autonomous Systems and Software Program and Nanyang Technological University (WASP/NTU) under grant M4082187 (4080), and Alibaba Group through Alibaba Innovative Research (AIR) Program and Alibaba-NTU Singapore Joint Research Institute (JRI), NTU-Singapore, Singapore Ministry of Education (MOE) Tier 1 (RG16/20). Zengxiang~Li is partially supported by RIE 2020 Advanced Manufacturing and Engineering (AME) Domain’s Core Funds-SERC Strategic Funds: Trusted Data Vault (Phase 1) (Grant No. A1918g0063). Yingbo~Liu is supported by National Natural Science Foundation of China (No. 11703010) (Corresponding author: Yingbo~Liu).

Yang Zhao, Jun Zhao, Linshan~Jiang, Rui~Tan and Dusit Niyato are with School of Computer Science and Engineering, Nanyang Technological University, Singapore, 639798. (Emails: s180049@e.ntu.edu.sg, junzhao@ntu.edu.sg, linshan001@e.ntu.edu.sg, tanrui@ntu.edu.sg, dniyato@ntu.edu.sg).}
\thanks{Zengxiang~Li is with the Institute of High Performance Computing (IHPC), A*STAR, Singapore. (Email: zengxiang$\textunderscore$li@outlook.com).
}
\thanks{Lingjuan~Lyu is with the Department of Computer Science, National University of Singapore. (Email: lingjuanlvsmile@gmail.com).
}
\thanks{Yingbo~Liu is with Big Data Research Institute of Yunnan Economy and Society, Yunnan University of Finance and Economics Kunming, China, 650201. (Email: liuyingbo@cnlab.net).}
}

\maketitle
\thispagestyle{fancy}
\pagestyle{fancy}
\lhead{This paper appears in IEEE Internet of Things Journal (IoT-J).  Please feel free to contact us for questions or remarks.}
\cfoot{\thepage}
\renewcommand{\headrulewidth}{0.4pt}
\renewcommand{\footrulewidth}{0pt}

\begin{abstract}
Home appliance manufacturers strive to obtain feedback from users to improve their products and services to build a smart home system. To help manufacturers develop a smart home system, we design a federated learning (FL) system leveraging a reputation mechanism to assist home appliance manufacturers to 
train a machine learning model based on customers' data. Then, manufacturers can predict customers' requirements and consumption behaviors in the future. The working flow of the system includes two stages: in the first stage, customers train the initial model provided by the manufacturer using both the mobile phone and the mobile edge computing (MEC) server. 
Customers collect data from various home appliances using phones, and then they download and train the initial model with their local data. 
After deriving local models, customers sign on their models and send them to the blockchain. In case customers or manufacturers are malicious, we use the blockchain to replace the centralized aggregator 
in the traditional FL system. Since records on the blockchain are untampered, malicious customers or manufacturers' activities are traceable. In the second stage, manufacturers select customers or organizations as miners for calculating the averaged model using received models from customers. 
By the end of the crowdsourcing task, one of the miners, who is selected as the temporary leader, uploads the model to the blockchain.
To protect customers' privacy and improve the test accuracy, we enforce differential privacy 
on the extracted features
and propose a new normalization technique. 
We experimentally demonstrate that
our normalization technique outperforms batch normalization when features are under differential privacy protection.
In addition, to attract more customers to participate in the crowdsourcing FL task, we design an incentive mechanism to award  participants.
\end{abstract}


\section{Introduction}
Internet of Things (IoT)-enabled smart home systems have gained great popularity in the last few years since they have an aim to increase the quality of life. A report by Statista~\cite{smarthomesta} estimates that by 2022, the smart home market size around the world will be $53.3$ billion. This smart home concept is mainly enabled by IoT devices, smart phone, modern wireless communications, cloud $\&$ edge computing, big data analytics, and artificial intelligence (AI). 
In particular, these advanced technologies enable manufacturers to maintain a seamless connection among their smart home appliances. %In the future, the IoT-enabled connected home appliances will be able to not only interact with each other but also with manufacturers to provide better services, and more devices will be added with smarter features. %However, with the passage of time, the smart home systems are depending on the collection of data. To help home appliance manufacturers improve their products to be smarter, we design a crowdsourcing federated learning (FL) system.
With the proliferation of smart home devices, tremendous data are generated. Federated learning (FL) enables analysts to analyze  and utilize  the locally generated data in a decentralized way without requiring  uploading data to a centralized  server; that is, the utility of data are well maintained despite data are preserved locally. To help home appliance manufacturers smartly and conveniently use data generated in customers' appliances, we design an FL-based system.
Our system considers home appliances of the same brand in a family as a unit, and a mobile phone is used to collect data from home appliances periodically and train the machine learning model locally~\cite{wang2018adaptive}.
Since mobile phones have limited computational power and battery life, we offload part of the training task to the edge computing sever. Then, the blockchain smart contract is leveraged to generate a global model by averaging the sum of locally trained models submitted by users. In this federated way, source data are supposed to maintain security and privacy. 

However, Melis~\emph{et~al.}~\cite{melis2018inference} demonstrated that gradient updates might leak significant information about customers' training data. Attackers can recover data from gradients uploaded by customers~\cite{hitaj2017deep}.  %Thus, customers may be unwilling to participate in our FL system. 
Besides, the federated approach for training the model is susceptible to model poisoning attacks~\cite{fung2018mitigating}.
%Therefore, customers who participate in federated learning and upload poisoning updates may negatively affect the global model. 
In addition, information leakage risks exist in the third party's mobile edge computing (MEC) server~\cite{zhang2020mdldroid}. To address aforementioned security and privacy issues, we adopt blockchain and differential privacy.  It is worth noting Apple is successfully applying differential privacy in FL to improve the privacy of its popular voice assistant service Siri~\cite{appledifferentialprivacy}. Specifically, manufacturers upload a preliminary model with initialized parameters. The model is available on the blockchain for customers to download and train with their local data. The blockchain assists the crowdsourcing requester (i.e. manufacturer) to audit whether there are malicious updates from customers. The traditional crowdsourcing system is hosted by a third party, which charges customers costly service fees, while our designed system uses blockchain to record crowdsourcing activities. Therefore, customers and the requester can save high service fees while keeping the crowdsourcing system functional. Due to the limitation of the block size, we propose to use the InterPlanetary File System (IPFS)~\cite{benet2014ipfs} as the distributed storage solution when the model size is large.

More specifically, customers extract their data's features in the mobile using the %designed CNN network 
deployed feature extractor and add noise with a formal privacy guarantee to perturb the extracted features in the first step. In the second step, customers train fully connected layers of the model with perturbed features in the MEC server. Moreover, we improve the traditional batch normalization by removing constraints of mean value and variance, while %lefting 
constraining the bound within $[-\sqrt{N-1}, \sqrt{N-1}]$, where $N$ denotes the batch size. After training, customers sign on hashes of encrypted models with their private keys and transmit locally trained models to the blockchain. Selected miners verify identities of senders, download models and calculate the average of all model parameters to obtain the global model. One miner, selected as the temporary leader, encrypts and uploads the global model to the blockchain. Furthermore, to motivate more customers to participate in the crowdsourcing task and reduce malicious and poisoning updates, we utilize a reputation-based crowdsourcing incentive mechanism, which rewards reliable customers and punish malicious customers correspondingly.

\textbf{Contributions.} The major contributions of this paper are summarized as follows:
\begin{itemize}
  \item First, a hierarchical crowdsourcing FL system is proposed to build the machine learning model to help home appliance manufacturers improve their service quality and optimize functionalities of home appliances.
  \item Second, we propose a new normalization technique which delivers a higher test accuracy than batch normalization, while preserving the privacy of the extracted features of each participant's data. Besides, by leveraging differential privacy, we prevent adversaries from exploiting the learned model to infer customers' sensitive information.
  \item Third, our blockchain-based system prevents malicious model updates by ensuring that all model updates are held accountable.
\end{itemize}

\textbf{Organization.}~The rest of the paper is organized as follows. In Section~\ref{sec-preliminary}, we explain the concepts of blockchain and differential privacy used in this paper. Section~\ref{sec-related} presents related work. We introduce our design of the system in Section~\ref{sec-system}. Section~\ref{sec-pro} shows the advantages and disadvantages of our designed system. Section~\ref{sec-experiment} presents the experimental results showing that our technique is working. Section~\ref{sec-discussion} discusses how we prevent information leakage using differential privacy technique in our designed system.  Then, we conclude the paper and identify future directions in Section~\ref{sec-conclusion}.

\textbf{Notations.} Notations used in the rest of the paper are summarized in Table~\ref{table:notation}.

\begin{table}[!h]
\caption{Summary of notations}
\centering
~~~~~~~~~~~~~\begin{tabular}{|l|l|lll}
\cline{1-2}
Symbol                  & Definition                                                                                                                                         &  &  &  \\ \cline{1-2}
$\epsilon$              & differential privacy budget                                                                                                                        &  &  &  \\ \cline{1-2}
$N$                     & batch size                                                                                                                                         &  &  &  \\ \cline{1-2}
$\mu$                   & mean value of normalized features                                                                                                                  &  &  &  \\ \cline{1-2}
$\sigma$                & variance  of  normalized  features                                                                                                                 &  &  &  \\ \cline{1-2}
$L_f$                     & length of feature                                                                                                                                  &  &  &  \\ \cline{1-2}
$W_f$                     & width of feature                                                                                                                                   &  &  &  \\ \cline{1-2}
$B$                     & each batch                                                                                                                                         &  &  &  \\ \cline{1-2}
$X_{i,j,k}$             & \begin{tabular}[c]{@{}l@{}}value at a position $\langle i , j \rangle$ for\\  the feature of image $k$\end{tabular}                                &  &  &  \\ \cline{1-2}
$\widetilde{X}_{i,j,k}$ & \begin{tabular}[c]{@{}l@{}}value at a position $\langle i , j \rangle$ for the \\ feature of image $k$ after batch normalization\end{tabular}      &  &  &  \\ \cline{1-2}
$\hat{X}_{i,j,k}$       & \begin{tabular}[c]{@{}l@{}}value at a position $\langle i , j \rangle$ for the \\ feature of image $k$ after our normalized technique\end{tabular} &  &  &  \\ \cline{1-2}
$s$                     & score                                                                                                                                              &  &  &  \\ \cline{1-2}
$R$                     & number of updates                                                                                                                                  &  &  &  \\ \cline{1-2}
$f$                     & Byzantine miners                                                                                                                                  &  &  &  \\ \cline{1-2}
$\Delta w$              & model update                                                                                                                                       &  &  &  \\ \cline{1-2}
$\gamma$              & reputation value                                                                                                                                   &  &  &  \\ \cline{1-2}
$\gamma^{Max}$        & maximal reputation value                                                                                                                           &  &  &  \\ \cline{1-2}
$h$                   & average reputation of the participating customers                                                                                                    &  &  &  \\ \cline{1-2}
$L$                   & low evaluation result                                                                                                    &  &  &  \\ \cline{1-2}
$H$                   & high evaluation result                                                                                                    &  &  &  \\ \cline{1-2}
\end{tabular}
\label{table:notation}
\end{table}

\section{Preliminaries}\label{sec-preliminary}
We organize this section on preliminaries as follows. In Section~\ref{subsection-blockchain}, we explain the concepts of blockchain and InterPlanetary File System. In Sections~\ref{subsection-dp} and~\ref{sec-FL}, we introduce the formal definitions of differential privacy and federated learning, respectively. 

\subsection{Blockchain and InterPlanetary File System (IPFS)}\label{subsection-blockchain}
The blockchain is a chain of blocks that contain the hash of the previous block, transaction information, and a timestamp. Blockchain originates from a bitcoin network as an append-only, distributed and decentralized ledger to record peer-to-peer transactions permanently and immutably. The IPFS is a peer-to-peer distributed file system that enables distributed computing devices to connect with the same file system. We implement the off-chain storage by using IPFS, and store hashes of data locations on the blockchain instead of actual files. The hash can be used to locate the exact file across the system.

\subsection{Differential Privacy}\label{subsection-dp}
Differential privacy (DP) guarantees the privacy and utility of a dataset with rigorous theoretical foundation~\cite{dwork2006calibrating, dwork2006our}. Intuitively, because the algorithm's output is perturbed by the noise, the absence or presence of one user's information in the database will not affect it much. By using the DP algorithm, analysts \mbox{cannot} derive confidential information when analyzing the algorithm's outputs. DP has received much interest in both the academia and the industry. For example, Apple's mobile operating system iOS utilizes DP algorithms in~\cite{tang2017privacy}. Google's Chrome browser implements a DP tool called RAPPOR for collecting customers' data~\cite{erlingsson2014rappor}. A smaller privacy parameter $\epsilon$ denotes the stronger privacy protection, but less utility of data as more randomness is added to the data. In contrast, a higher privacy parameter $\epsilon$ denotes the weaker privacy protection but ensuring better utility of the data. The formal definition for differential privacy is as follows :

\begin{defn} (Differential Privacy.) A randomized mechanism $\mathcal{M}$ provides $(\epsilon, \delta)$-differential privacy if any two neighbouring datasets $D$ and $D'$ (i.e., $D$ and $D'$ differ in  at most one record), $\mathcal{M}$ guarantees that 
\begin{align}
   \textup{Pr} [\mathcal{M}(D) \in Y] \leq e^{\epsilon} \textup{Pr}[\mathcal{M}(D') \in Y] + \delta \nonumber,
\end{align}
where $\textup{Pr}[\cdot]$ denotes the probability, and the probability space is over the coin flips of the randomized mechanism $\mathcal{M}$. $Y$ iterates through all subsets of the output range of mechanism $\mathcal{M}$. When $\delta = 0$, the mechanism $\mathcal{M}$ becomes $\epsilon$-differentially private.
\end{defn}

The \textbf{Laplace mechanism} of~\cite{dwork2006calibrating} can be used to ensure differential privacy by adding independent \mbox{zero-mean} Laplace noise with scale $\lambda$ to each dimension of the output. Specifically, $\lambda$ equals $\Delta/\epsilon$, where $\Delta$ is the $\ell_1$-norm sensitivity of the query $Q$ and it measures the maximum change of the true query output over neighboring databases.

The \textbf{post-processing} property~\cite{dwork2006calibrating} of differential privacy means that a data analyst cannot compute the function of the output of a differentially private algorithm and reduce its privacy guarantee without obtaining additional knowledge about the private data. Hence, in our design, although noise is added to an intermediate layer of the neural network, the post-processing property ensures that the final trained model also satisfies differential privacy.

\subsection{Federated Learning}\label{sec-FL}
Traditional machine learning algorithms are conducted in a centralized data center where data owners upload their data. However, data are privacy sensitive, and data owners are unwilling to share; thus, collecting data is a tough and verbose task that hinges the progress of machine learning. To avoid the data deficiency problem as well as maintain the  machine learning model's accuracy and performance, a decentralized approach of conducting machine learning, federated learning (FL), is proposed~\cite{mcmahan2017communication}; that is, data are distributed and scattered among different users, and no such a single node stores the whole dataset. The workflow of FL is that each user trains a local machine learning model using her dataset and uploads it to the centralized model for summarizing and averaging. Then, a global model is achieved in the centralized server. Thus, FL prevents a single point of failure effectively. FL is similar to the traditional distributed machine learning~\cite{konevcny2016federated}, but assumptions on local datasets are different. More specifically, the traditional distributed learning aims at optimizing the parallel computing power, but data are IID among different parties, while FL focuses on the heterogeneous local datasets, meaning that training data can be distributed, non-IID, and unbalanced among various participants. That is, each participant $i$ trains the same model with her local dataset, and the goal is to obtain a global model with the minimized averaged sum of loss functions among all participants.

\begin{figure*}[!t]
\centering
\includegraphics[scale=0.5]{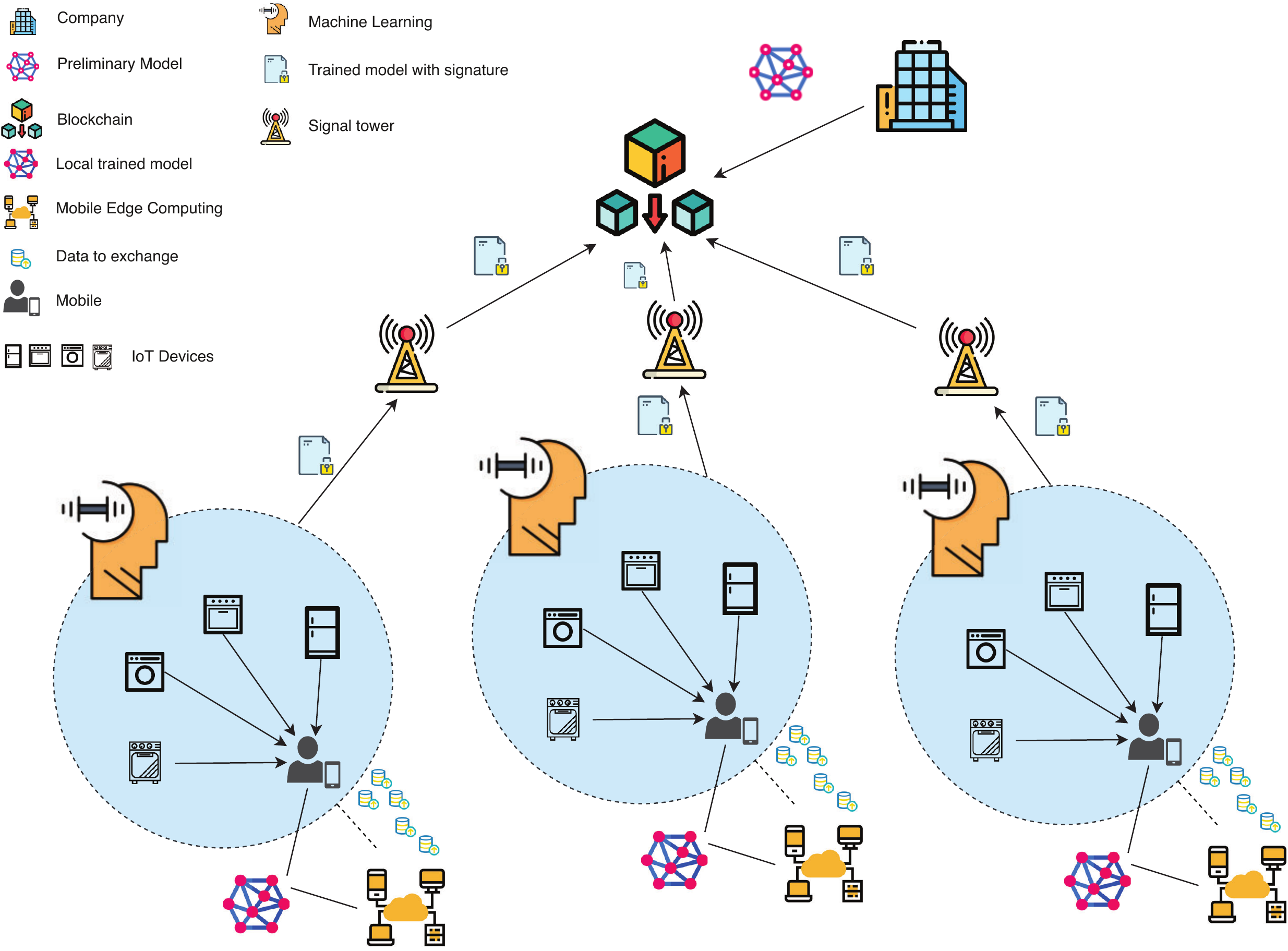}
\caption{An overview of our system.}
\label{fig:accuracy}
\end{figure*}

\section{Related Work}\label{sec-related}
Blockchain and federated learning (FL) techniques have been widely used in training a neural network with distributed data~\cite{qu2019proof, lu2019blockchain, ramanan2019baffle, kim2019blockchained, yin2020fdc, awan2019poster, weng2019deepchain,lyu2019towards,ZhaoWZZC19,ZhaoHWJSLH20}. For example, Weng~\emph{et~al.}~\cite{weng2019deepchain} proposed a system called DeepChain for the collaborative learning. But they did not offload the training task to the edge server, and they did not  propose to use differential privacy to protect the  privacy  of  model parameters. Awan~\emph{et~al.}~\cite{awan2019poster} proposed a blockchain-based privacy-preserving FL framework, which secured the model update using blockchain's immutability and decentralized trust properties.
Li~\emph{et~al.}~\cite{li2018crowdbc} designed a decentralized framework based on blockchain for crowdsourcing tasks, which enabled them to do crowdsourcing tasks without a centralized server. Lu~\emph{et~al.}~\cite{lu2019blockchain} proposed to leverage blockchain, FL, and differential privacy for data sharing. However, they %proposed adding 
directly added differential privacy noise to the original data instead of the gradients, which may affect the accuracy seriously.
Lyu~\emph{et~al.}~\cite{lyu2019towards} made the first-ever investigation on the federated fairness in a blockchain-assisted decentralized deep learning framework, and designed a local credibility mutual evaluation mechanism to enforce fairness. They also developed a scheme for encryption  to ensure  privacy and accuracy.

Moreover, FL %has attracted much
has attracted substantial attention recently~\cite{yang2019federated, lim2019federated, li2019federated, nishio2019client,lyu2020threats}, %and some studies discussed privacy issues in federated learning
and one of the most important issues in FL is privacy protection, which is explored in~\cite{liu2019enhancing, hao2019efficient, dolui2019poster, nasr2019comprehensive, wang2019beyond}. Li~\emph{et~al.}~\cite{liu2019enhancing} considered the privacy issue during sharing model updates in FL. They proposed to leverage the sketch algorithms to build the sketching-based FL, which provides privacy guarantees while maintaining the accuracy. Hao~\emph{et~al.}~\cite{hao2019efficient} proposed a privacy-enhanced FL scheme to solve the privacy issue in FL. Their scheme helps to achieve efficient and privacy-preserving FL. Dolui~\emph{et~al.}~\cite{dolui2019poster} applied FL paradigms to recommender systems and matrix factorization, which guarantees the recommender systems' functionality and privacy. Nasr~\emph{et~al.}~\cite{nasr2019comprehensive} performed a comprehensive privacy analysis with white-box inference attacks. Wang~\emph{et~al.}~\cite{wang2019beyond} proposed a framework incorporating  generative adversarial network with a multitask discriminator to solve the user-level privacy leakage in FL against attacks from a malicious server.

Furthermore, there are many studies focusing on privacy-preserving crowdsourcing~\cite{wu2020crowdprivacy, liang2020enabling}, and leveraging fog computing or edge computing to improve the performance as they have gained popularity~\cite{he2020privbus, zhang2020fog, zhao2020p3,xu2019blockchain, shi2016edge, shi2016promise}.  For example, Wu~\emph{et~al.}~\cite{wu2020crowdprivacy} proposed two generic models for quantifying mobile users' privacy and data utility in crowdsourced location-based services, respectively. He~\emph{et~al.}~\cite{he2020privbus}~designed a privacy model for the crowdsourced bus service, which takes advantage of the computational power of the fog computing. However, their models are applicable only to the traditional crowdsourcing approach (i.e., customers transmit data to a centralized server) without considering the FL crowdsourcing tasks which leverage locally trained models. Zhao~\emph{et~al.}~\cite{zhao2020p3} presented with a privacy-preserving mechanism to prevent the poisoning attack to the mobile edge computing. However, users need to offload data to the MEC server in their system which may leak the privacy, instead, we propose that users retain their data locally. %We use the poisoning attacks protocol, Multi-Krum, as part of the incentive mechanism.

In addition, a few studies have combined deep learning or FL with edge computing~\cite{lyu2019fog,jiangdifferentially, wang2019adaptive, wang2019edge}.
Lyu~\emph{et~al.}~\cite{lyu2019fog} proposed a privacy-preserving deep learning framework, where a two-level protection mechanism, including \textit{Random Projection} and \textit{Differentially Private SGD}, is leveraged to protect the data privacy. Jiang~\emph{et~al.}~\cite{jiangdifferentially} designed a collaborative training method to protect features' privacy. In detail, the feature extraction is done locally in the devices such as smartphones while the classification is executed in the cloud service. However, they did not use the FL to protect the privacy of training data, and they did not propose a normalization technique to improve the test accuracy. Wang~\emph{et~al.}~\cite{wang2019adaptive, wang2019edge} proposed control algorithms to solve the problem of low resources in IoT devices while participating in the federated learning.

\section{System Design}\label{sec-system}

In this section, we present the system which we design  for smart home appliance manufacturers who are interested in building a machine learning model using data from customers' home appliances to analyze customers' habit and improve their service and products.

\subsection{System Overview}

Figure~\ref{fig:accuracy} shows an overview of our system architecture. The system consists of three primary components: manufacturers, customers and blockchain. Specifically, manufacturers raise a request for a crowdsourcing FL task. Then, customers who are interested in the crowdsourcing FL tasks submit their trained models to the blockchain. Finally, the blockchain serves as the centralized  server to gather customers' models, and a selected miner calculates and generates the global FL model for home appliance manufacturers. In the following, we will introduce each component in detail.

\textbf{\textit{Manufacturers.}} %We consider building a machine learning model to help manufacturers to predict customers' consumption behaviours and improve home appliances. 
Manufacturers raise a request to build a machine learning model to predict customers' consumption behaviours and improve home appliances, which is a crowdsourcing FL task. Customers who have home appliances can participate in the FL task. %The crowdsourcing system implemented using the blockchain to record the progress of the task. 
% \subsubsection{Manufacturers put a preliminary model in the blockchain}
To facilitate the progress of FL, we use the blockchain to store the initial model with randomly selected parameters. Otherwise, manufacturers need to send the model to everyone or save it in a third party's cloud storage. In addition, neither manufacturers nor customers can deny recorded contributions or activities.  Eventually, manufacturers will learn a machine learning model as more and more customers participate in the crowdsourcing FL task.

\textbf{\textit{Customers.}}\label{subsec:customers} Customers who have home appliances satisfying crowdsourcing requirements can apply for participating in the FL task. %Based on the incentive mechanism, customers will be rewarded with coins and reputation based on their contributions. %Large home appliance manufacturers always produce a variety of home appliances. 
However, since home appliances are  equipped with heterogeneous storage and computational powers, it is difficult to enable each IoT device to train the deep model. To address this issue, we adopt the partitioned deep model training approach~\cite{jiangdifferentially, mao2018learning}. Specifically, we use a mobile phone to collect data from home appliances and extract features. To preserve privacy, we add $\epsilon$-DP noise to features. Then, customers continue training fully connected layers in the MEC server. To be specific, we clarify the customers' responsibilities in four detailed steps as follows.

\textit{ Step 1: Customers download the initial model from the blockchain.} Customers who are willing to  participate  in the  FL task  check and download the initial model which is uploaded by the manufactures and available on the blockchain.

\textit{ Step 2: Customers extract features on the mobile.} The mobile phone collects all participating home appliances' data periodically. Then, customers can start training the model using collected data. Since the MEC server is provided by a third-party, it may leak information. Therefore, we divide the local training process into two phrases: the mobile training and the MEC server training. Because perturbing original data directly may compromise the model's accuracy, we treat the convolutional neural network (CNN) layers as the feature extractor to extract features from the original data in the mobile. Then, we add $\epsilon$-DP noise to features before offloading them to the fully connected layers in the MEC.

\textit{ Step 3: Customers train fully connected layers in the mobile edge computing server.} The mobile sends the privacy-preserving features and original labels to the mobile edge computing server, so that the server helps train the fully connected layers. The training loss is returned to the mobile to update the front layers.

\textit{ Step 4: Customers upload models to the blockchain.}
After training the model, customers sign on hashes of  models with their private keys, and then they upload models to the blockchain via smartphones. However, if miners determine that the signature is invalid, the transaction fails because it is possible that an adversary is attempting to attack the learning process using faked data. After miners confirm the transaction, customers can use the transaction history as an invoice to claim reward and reputation. Section~\ref{sec-incentive-mechanism} shows the detail of reputation calculation. By using the immutable property of the blockchain, both manufacturers and customers cannot deny transactions stored on the blockchain.

\textbf{\textit{Blockchain.}} A consortium blockchain is used in our crowdsourcing system to store machine learning models permanently. The consensus protocol is the Algorand which is based on proof of stake (PoS)  and  byzantine fault tolerance (BFT)~\cite{wang2019survey,gilad2017algorand}. Algorand relies on BFT algorithms for committing transactions. The following steps are required to reach the consensus: (1) Miners compete for the leader. The ratio of a miner's stake (i.e., coins) to all tokens determines the probability for the miner to be selected. Subsequently, by hashing the output of random function with the identities of nodes which are specified by their stake, the order of the block proposals is obtained. Thus, a miner with more stakes will gain a higher chance to become a leader. (2) Committee members verify the block generated by the selected leader. When more than $2/3$ of the committee members sign and agree on the leader's block, the new block gets admitted.  (3) Committee members execute  the gossip protocol to broadcast the new block to neighbours to arrive at a consensus in blockchain.

In our case, the workflow starts with a manufacturer uploading an initial model to the blockchain. Then, customers can send requests to obtain that model. After training models locally, customers upload their locally trained models to the blockchain. Because of the limitation of the block size, we propose to use IPFS as the off-chain storage. Then, customers upload their models to the IPFS, and a hash will be sent to the blockchain as a transaction. The hash can be used to retrieve the actual data from IPFS. %Miners are responsible for confirming transactions and calculating the averaged model parameters.
The leader and miners are responsible for confirming transactions and calculating the averaged model parameters to obtain a global model. Miners' results are mainly used for verifying the leader's result.  %After customers upload their locally trained models to the blockchain, miners verify that uploaded models' signatures are valid and confirm the transactions. 
After all customers upload their trained models, the miners download them and start calculating the averaged model parameters. Then, one of miners is selected as the leader to upload the global model to the blockchain. We will explain the process in detail as follows:

\textit{\ding{172} Miners verify the validity of the uploaded model.} When a customer uploads a model or the hash of the model to the blockchain, a miner checks the digital signature of the uploaded file. If the signature is valid, then the miner confirms that the update is from the legal participant and puts the transaction in the transaction pool. Subsequently, selected miners constitute a committee to verify all transactions in the pool using Multi-KRUM~\cite{blanchard2017machine, shayan2018biscotti},  and accept legitimate updates. After verifying the validity of the uploaded model, the leader selected from miners will generate a new block containing the uploaded file. 

\textit{\ding{173} A selected leader updates the model.}
A leader is selected from a group of miners to update the model. Miners compete for updating parameters to get the reward. Algorand uses the Verifiable Random Functions (VRF) as a local and non-interactive way to select a subset of users as the leader candidates who form a committee (weighed by their coins) and determine their priorities. A leader candidate with the highest priority will become the leader to update the model parameters. As each user is weighted by their coins, one unit of coin can be regarded as a sub-user. A user with $m$ coins has $m$ ``sub-users''. Let $\tau$ be the expected number of sub-users that the system desires to choose, and $M$ be the total mount of coins of all users. Then the probability $p$ of any coin being chosen can be set as $\tau / M$. For a user $u$ with $m$ units of currency, it will first use its secret key to generate a hash and proof via VRF. The interval $[0, 1]$ is divided into $m+1$ sub-intervals, so that the number $j$ of selected sub-users for this user is determined by which sub-interval $hash/2^{hashlen}$ falls in ($hashlen$ denotes the length of $hash$); i.e., $j$ satisfies $hash/2^{hashlen} \in [ \sum_{k=0}^j \binom{m}{k} p^k (1-p)^{m-k}, \sum_{k=0}^{j+1} \binom{m}{k} p^k (1-p)^{m-k} )$ (if $hash/2^{hashlen} =1$, then $j = m$). Other users can use the proof to check that user $u$ indeed has $j$ sub-users selected. The number of selected sub-users is each user's priority. The user with the highest priority will become the leader. The selected leader is responsible for aggregating models submitted by customers and uploading the global model to the blockchain.

\subsection{Incentive mechanism}\label{sec-incentive-mechanism}
To attract more customers to contribute to building the FL model, we design an incentive mechanism. Because data in home appliances contain customers' confidential information and training consumes computing resources, some customers are unwilling to participate in training the FL model. However, with an incentive mechanism, customers will be rewarded based on their contributions. Then, customers may trade for services, such as the maintenance and upgrade services for appliances, provided by manufacturers using rewards. Specifically, by combining the Multi-KRUM~\cite{blanchard2017machine, shayan2018biscotti} and the reputation-based incentive protocols~\cite{zhang2012reputation}, an incentive mechanism is designed to prevent the poisoning attack as well as reward contributors properly.

That is, after the local model is uploaded, verifiers calculate the reputation using Multi-KRUM algorithm and eliminate unsatisfied updates.  The verifiers, selected based on the VRF~\cite{gilad2017algorand} from miners, will remove malicious updates  by executing Multi-KRUM algorithm on updates in the received pool and accept the top majority of the updates received every global epoch. The verifier will add up Euclidean distances of each customer $i$'s update to the closest $R-f-2$ updates and denote the sum as each customer $i$'s score $s(i)$. $R$ means the 
number of updates, and $f$ means the number of Byzantine customers.  $\Delta w$ means the model update. It is given by
\begin{align}
    s(i) = \sum_{i \rightarrow j} \|\Delta w_i  - \Delta w_j   \|^2,\label{eq:distance}
\end{align}
where $i \rightarrow j$ denotes the  fact that $\Delta w_j$ belongs to the $R-f-2$ closest updates to $\Delta w_i$. The $R-f$ customers who obtain the lowest scores will be chosen while rejecting the rest.

The value of the reward is proportional to the customer's  reputation. If a customer's update is accepted by verifiers, the value of reputation increases by $1$; otherwise, it decreases by $1$. Each participant is assigned with an initial reputation value $\gamma$, and $\gamma$ is an integer selected from the \textit{set}$(0, 1, \cdots, \gamma^{Max})$, where $\gamma^{Max}$ denotes the highest reputation. $h$ denotes the average reputation of the whole customers. If a miner verifies a solution is correct and provides a positive evaluation, the reputation of the participant will be increased and recorded in the blockchain. Let $a$ denote the evaluation function's output. $a = H$ denotes a high evaluation result while $a = L$ denotes a low evaluation result. Therefore, the update rule of the reputation $\gamma$ is as follows:

\begin{align}
    \gamma = \begin{cases} \min(\gamma^{Max}, \gamma + 1), &\textup{if~} a = H~\textup{and}~\gamma \geq h \\
    \gamma -1, &\textup{if~} a = L~\textup{and}~\gamma \geq h + 1 \\
    0, &\textup{if~} a = L~\textup{and}~\gamma = h \\
    \gamma +1, &\textup{if~} \gamma < h \\
    \end{cases}
\end{align}
where $h$ denotes the threshold of the selected social strategy, which is a method of using social norms (i.e., Multi-KRUM) to control customers' behaviours~\cite{zhang2012reputation}. If a customer's reputation is $h$ and she receives an $L$ feedback after evaluation, her reputation will fall to $0$. The status of customers' reputation is recorded by the blockchain.

\subsection{Normalization Technique}

To protect the privacy of users' update, we perturb  extracted features in the normialization layer. Now, we present the improvement for the normalization technique proposed in~\cite{jiangdifferentially}. Although the CNN has many channels, our analysis below focuses on one channel only for simplicity. For this channel, suppose the output of the convolutional
layers has dimension $L_f \times W_f$. Let the value at a position $\langle i , j \rangle$ for the feature of image $k$ be $X_{i,j,k}$. Given $i$ and $j$, Jiang~\emph{et~al.}~\cite{jiangdifferentially} adopt the batch normalization which transforms $X_{i,j,k}$ to $\widetilde{X}_{i,j,k}$, so that for each batch $B$, the values $\widetilde{X}_{i,j,k}$ for $k \in B$ have a  mean of $0$ and a variance of $1$; i.e., 
$$\frac{1}{|B|}\sum_{k \in B}\widetilde{X}_{i,j,k} = 0,$$ while $$ \frac{1}{|B|} \sum_{k \in B}(\widetilde{X}_{i,j,k})^2  =   1.$$
From and $|B|=N$ and the Cauchy--Schwarz inequality,~\cite{jiangdifferentially} bounds $$\widetilde{X}_{i,j,k} \in [- \sqrt{N-1} , \sqrt{N-1}]$$ for any $i,j,k$, so that if one value in the feature $$\big\{X_{i,j,k}~|~ i \in \{1,2,\ldots,L_f\} ~\textup{and}~ j \in \{1,2,\ldots,W_f\}\big\}$$ of image $k$ varies, the sensitivity of $$\big\{\widetilde{X}_{i,j,k}~|~ i \in \{1,2,\ldots,L_f\} ~\textup{and}~ j \in \{1,2,\ldots,W_f\}\big\}$$ is at most $ 2 \sqrt{N-1}$. 

Then, according to Laplace mechanism~\cite{dwork2006calibrating}, the independent \mbox{zero-mean} Laplace noise with scale ${2 \sqrt{N-1} }\big/{\epsilon}$ is added to each $\widetilde{X}_{i,j,k}$ for $i \in \{1,2,\ldots,L_f\} $ and $ j \in \{1,2,\ldots,W_f\}$ to protect ${X}_{i,j,k}$ under $\epsilon$-differential privacy. In our approach, we normalize ${X}_{i,j,k}$ for $i \in \{1,2,\ldots,L_f\} $ and $ j \in \{1,2,\ldots,W_f\}$ as $${\hat{X}}_{i,j,k} \in [- \sqrt{N-1} , \sqrt{N-1}],$$ so that if one value in the feature $$\big\{X_{i,j,k}~|~ i \in \{1,2,\ldots,L_f\} ~\textup{and}~ j \in \{1,2,\ldots,W_f\}\big\}$$ of image $k$ varies, the sensitivity of $$\big\{\hat{X}_{i,j,k}~|~ i \in \{1,2,\ldots,L_f\} ~\textup{and}~ j \in \{1,2,\ldots,W_f\}\big\}$$ is $2 \sqrt{N-1}$. Then, based on Laplace mechanism \cite{dwork2006calibrating}, the independent \mbox{zero-mean} Laplace noise with scale ${2 \sqrt{N-1} }\big/{\epsilon}$ is added to each $\hat{X}_{i,j,k}$ for $i \in \{1,2,\ldots,L_f\}  ~\textup{and}~  j \in \{1,2,\ldots,W_f\}$ to protect ${X}_{i,j,k}$ under $\epsilon$-differential privacy. From the above discussions, batch normalization of~\cite{jiangdifferentially} enforces not only $$\widetilde{X}_{i,j,k} \in [- \sqrt{N-1} , \sqrt{N-1}]$$ but also the mean is $$ \frac{1}{|B|}\sum_{k \in B}\widetilde{X}_{i,j,k} = 0$$ and the variance is $$ \frac{1}{|B|} \sum_{k \in B}(\widetilde{X}_{i,j,k})^2  =  1,$$ while our normalization technique requires only $$\hat{X}_{i,j,k} \in [- \sqrt{N-1} , \sqrt{N-1}]$$ without any constraints on the mean and variance. Experiments to be presented in Section~\ref{sec-experiment} show that our normalization technique significantly improves the learning accuracy over that of~\cite{jiangdifferentially}.

Next, we explain why our normalization technique outperforms the batch normalization. Both Jiang~\emph{et~al.}'s solution~\cite{jiangdifferentially} and our solution add the same zero-mean Laplace noise to normalized layer inputs. When using batch normalization, the mean of features $\mu = 0$ and the variance $\sigma = 1$. For  ease of explanation, below we use a Gaussian distribution as an example for the distribution of the features since Gaussian distributions appear in many real-world applications. Note that the actual distribution of the features may not follow Gaussian. According to the three-sigma rule of Gaussian distribution~\cite{pukelsheim1994three}, about $99.73\%$ values lie within three standard deviations of the mean. Similarly, most feature values after batch normalization lie in $[-3\sigma, 3\sigma]$ which is $[-3, 3]$ instead of $[-\sqrt{N-1}, \sqrt{N-1}]$. In contrast, feature values lie more evenly in $[-\sqrt{N-1}, \sqrt{N-1}]$ when using our normalization technique. Thus, features have smaller magnitudes when using batch normalization than using our normalization technique. Hence, %if added to 
when the same amount of Laplace noise is added, feature values using batch normalization will be perturbed more easily than using our normalization technique. For example, when the batch size $N = 64$ and scale of Laplace distribution is $2\sqrt{N-1}/\epsilon$, we calculate privacy parameter thresholds for feature values (after batch normalization or our normalization technique) to be ``overwhelmed" by the noise  as follows.

In the case of batch normalization, we have
\begin{align}
    \frac{2\sqrt{N-1}}{\epsilon} &\gg 3\sigma, \nonumber \\
  \Longrightarrow  \frac{2\sqrt{N-1}}{\epsilon} &\gg 3, \nonumber \\
  \Longrightarrow   \epsilon \ll \frac{16}{3} &\approx 5.33. \nonumber
\end{align}
Thus, true feature values will be seriously perturbed by noise when the privacy parameter $\epsilon \ll 5.33$ using batch normalization. However, when we use our normalization technique, we obtain that
\begin{align}
    \frac{2\sqrt{N-1}}{\epsilon} &\gg \sqrt{N-1}, \nonumber \\
   \Longrightarrow  \epsilon &\ll 2. \nonumber
\end{align}
Hence, when the privacy parameter $\epsilon \ll 2$, the true value will be overwhelmed by the noise. The larger privacy parameter means the less noise, so that feature values using batch normalization are more vulnerable. Thus, our above example implies that features will be perturbed more seriously when using batch normalization than our normalization technique. Summarizing above, the trained model with our normalization technique will achieve a higher test accuracy than trained using the batch normalization.

\section{Pros and Cons of our framework}\label{sec-pro}
We discuss advantages and disadvantages of our framework in this section.
\subsection{Privacy and Security}
Our system leverages differential privacy technique to protect the privacy of the extracted features.
Thus, the system keeps the participating customers' data confidential. 
Furthermore, the trained model is encrypted and signed by the sender to prevent the attackers and imposters from stealing the model or deriving original data through reverse-engineering.

\subsection{Delay in Crowdsourcing}
Assume there is a large number of customers, and the system highly depends on customers' training results to obtain the predictive model in one global epoch. Unlike other crowdsourcing jobs, manufacturers in our system prefers customers to follow their lifestyle instead of rushing to finish the job to obtain the real status. As a result, customers who seldom use devices may postpone the overall crowdsourcing progress. This problem can be mitigated by using the incentive mechanisms.  Yu~\emph{et~al.}~\cite{yu2020fairness} designed a queue to store customers who submitted their models in order. Thus, customers who submit their locally trained models early will be rewarded to encourage people to submit their updates earlier.

\section{Experiments} \label{sec-experiment}

\begin{figure*}[!t]
\centering
\includegraphics[width=\linewidth]{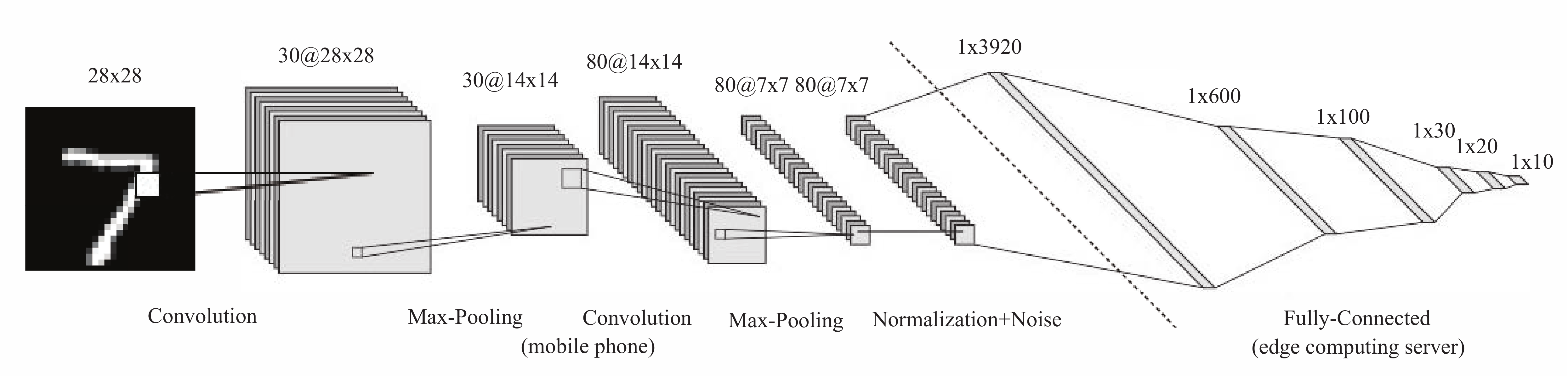}\vspace{-5pt}
\caption{%The CNN structure used in our experiments. The normalization layer and the noise followed are used to achieve differential privacy (DP) guarantee.
The neural network used in experiments.\vspace{-2pt}}
\label{network-structure}
\end{figure*}

To validate the effectiveness of our designed FL with differential privacy approach, we conduct experiments %about our designed federated learning with differential privacy method using
on the MNIST handwritten image dataset~\cite{lecun2010mnist}.
\subsection{Experiment Setup}
%We use the public database MNIST which 
The MNIST dataset includes 60,000 training image samples and 10,000 test image samples. Each sample is a $28 \times 28$ gray scale image showing a handwritten number within $0$ to $9$. In addition, the  MNIST is a standard dataset employed for testing machine learning algorithms. It gives moderate and typical complexity faced by IoT applications. Therefore, we leverage the MNIST dataset which has been used for testing the performance of the IoT system by~\cite{xu2020lightweight, zheng2019challenges, jiangdifferentially, mills2019communication, liu2019fitcnn, scheidegger2019constrained, kumagai2019transfer}. Our designed CNN network includes hidden layers that are responsible for feature extraction and fully connected layers for classification. We have two convolutional layers with $30$ and $80$ channels, respectively. After each convolutional layer, we deploy a max-pooling layer to reduce spatial dimensions of the convolutional layers' output. Therefore, max-pooling layers accelerate the learning speed of the neural network. Normalization is used after all non-linear layers, i.e., convolutional layers. The normalization layer enables the computation of sensitivity in differential privacy to determine the amount of noise to add, speeds up the learning rate, and regularizes gradients from distraction to outliers. Then, we apply $\epsilon$-DP noise to perturb the output of normalization layers to preserve the privacy of the extracted features. 

The perturbed features serve as inputs of fully connected layers for classification in the MEC server. In our designed model, fully connected layers include four hidden layers. The dimension decreases from $3920$ to the dimension of the label which is $10$. Finally, there is a softmax layer to predict label and compute loss. The architecture of CNN is shown in Figure~\ref{network-structure}. We simulate FL by constructing the model using the averaged parameters of multiple locally trained model parameters. 

In our experiment, we set the hyperparameters of CNN as follows. The learning rate is $0.01$, and the batch size $N$ is $64$. Then, we set the range of privacy parameter $\epsilon$ to be $[1,10]$.  The default number of global epochs is $2$ and the default number of local epochs is $40$. We use ten participants in the experiment. Before training, we separate the training image dataset into equally ten parts, meaning that each participant gets $6000$ training images randomly. We normalize each dimension of the feature to the interval $[-\sqrt{N-1}, \sqrt{N-1}]$ for $N$ denoting the batch size, so that the sensitivity of the normalized feature vector when one dimension of the  feature changes is $2 \sqrt{N-1}$. Then, according to Laplace mechanism~\cite{dwork2006calibrating}, the independent \mbox{zero-mean} Laplace noise with scale ${2 \sqrt{N-1} }\big/{\epsilon}$ is added to each dimension of the normalized features to protect features under $\epsilon$-differential privacy. Default $\epsilon = 2$.

\begin{figure}[!t]
\centering
\includegraphics[width=\linewidth]{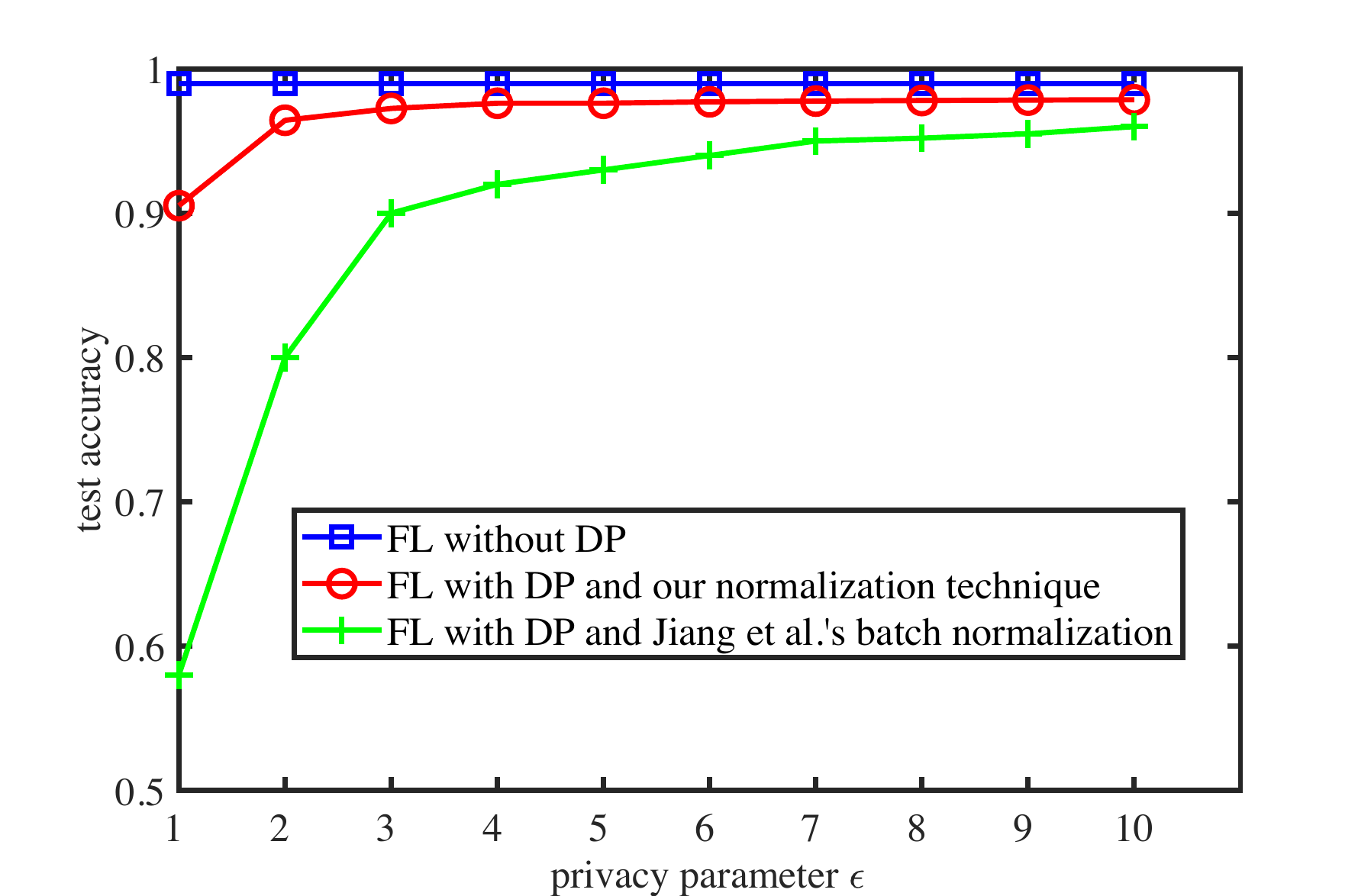}\vspace{-5pt}
\caption{
%Comparing the test accuracies between federated learning (FL) without differential privacy (DP) and different DP-aware FL algorithms, including DP-aware FL using our normalization technique, and DP-aware FL using Jiang~\emph{et~al.}'s~\cite{jiangdifferentially} batch normalization.
Impacts of normalization techniques on the test accuracy.\vspace{-12pt}}
\label{fig:FL-dp-no}
\end{figure}

\subsection{Experimental Results}

\begin{figure}[!t]
\centering
\includegraphics[width=\linewidth]{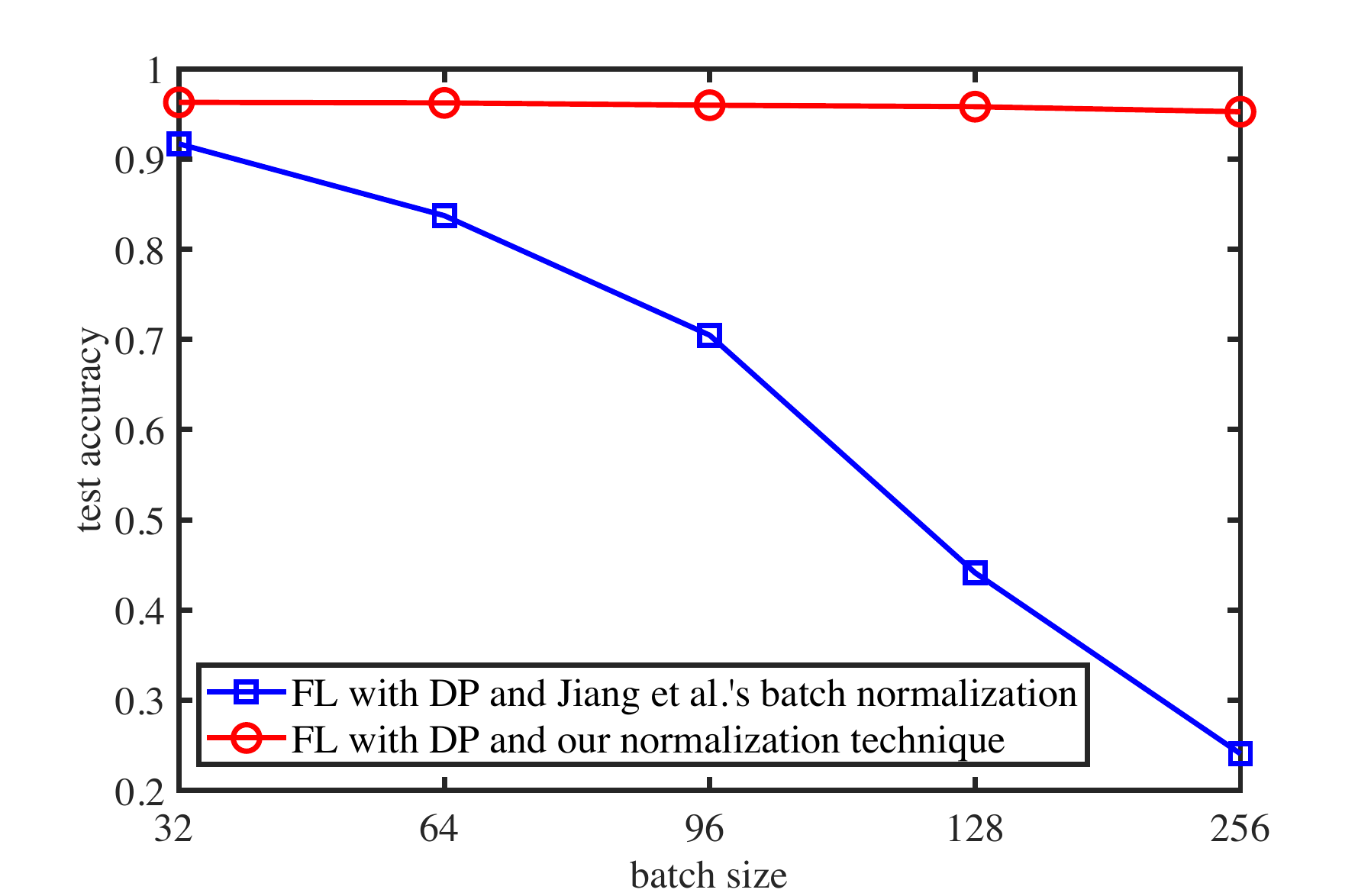}\vspace{-5pt}
\caption{%Impact of the batch size on the test accuracy of FL model with $\epsilon$-differential privacy for $\epsilon$ and local epochs being $2$ and $40$ and the global epoch $= 1$. 
Impact of the batch size on the test accuracy of the FL model protected with DP ($\epsilon = 2$).\vspace{-12pt}}
\label{fig:FL-bz-256}
\end{figure}

\begin{figure}[!t]
\centering
\includegraphics[width=\linewidth]{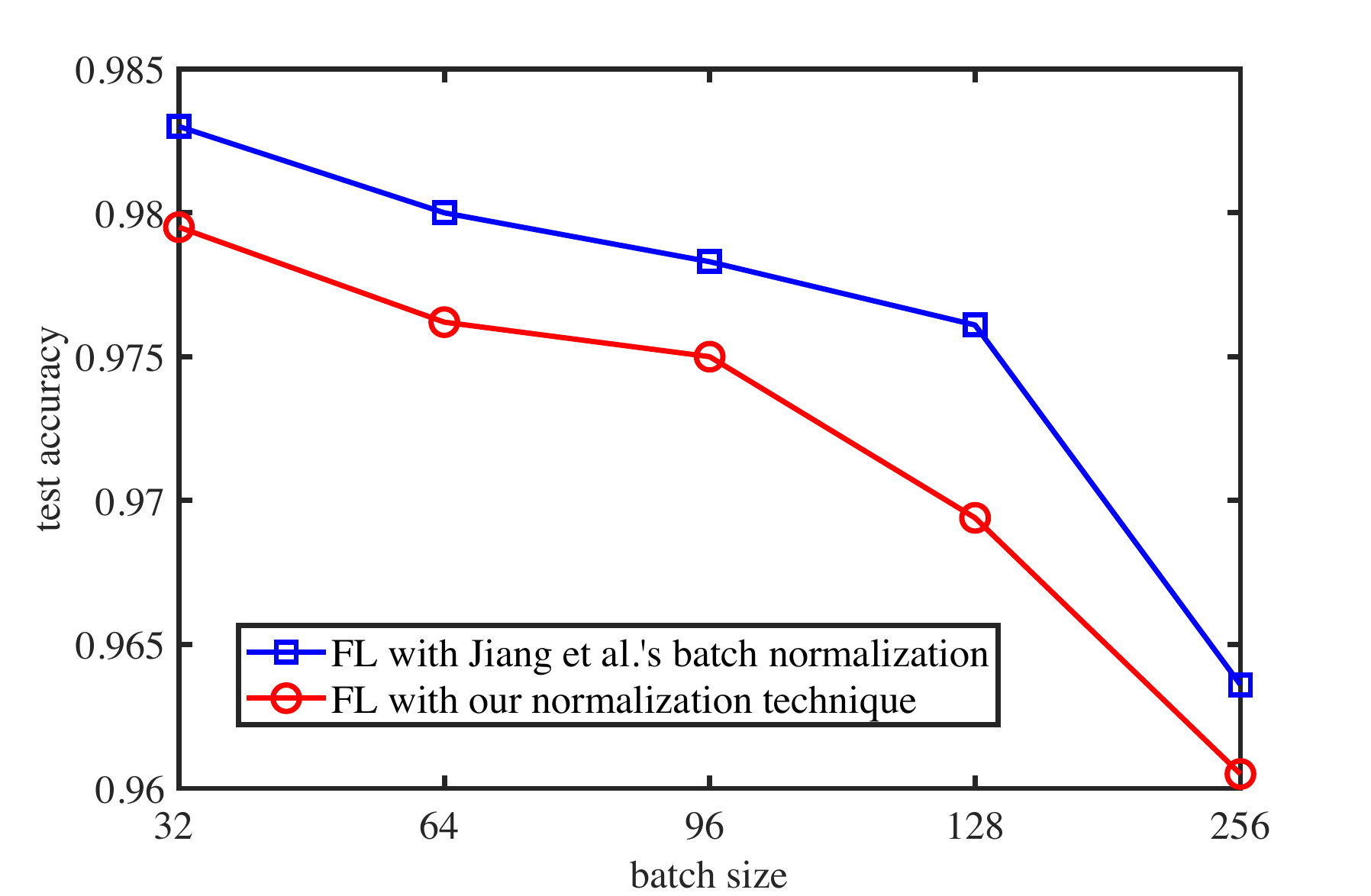}\vspace{-5pt}
%\caption{Comparing test accuracies between FL using our normalization technique and FL using Jiang~\emph{et~al.}'s~\cite{jiangdifferentially} batch normalization when the global epoch is $1$, $\epsilon$-differential privacy parameter $\epsilon = 2$ and the number of local epochs is $40$.\vspace{-2pt} \vspace{-2pt}}
\caption{Impact of the batch size on the test accuracy of the FL model using our normalization technique without DP protection.\vspace{-12pt}}
\label{fig:FL-no-dp}
\end{figure}

\begin{figure}[!t]
\centering
\includegraphics[width=\linewidth]{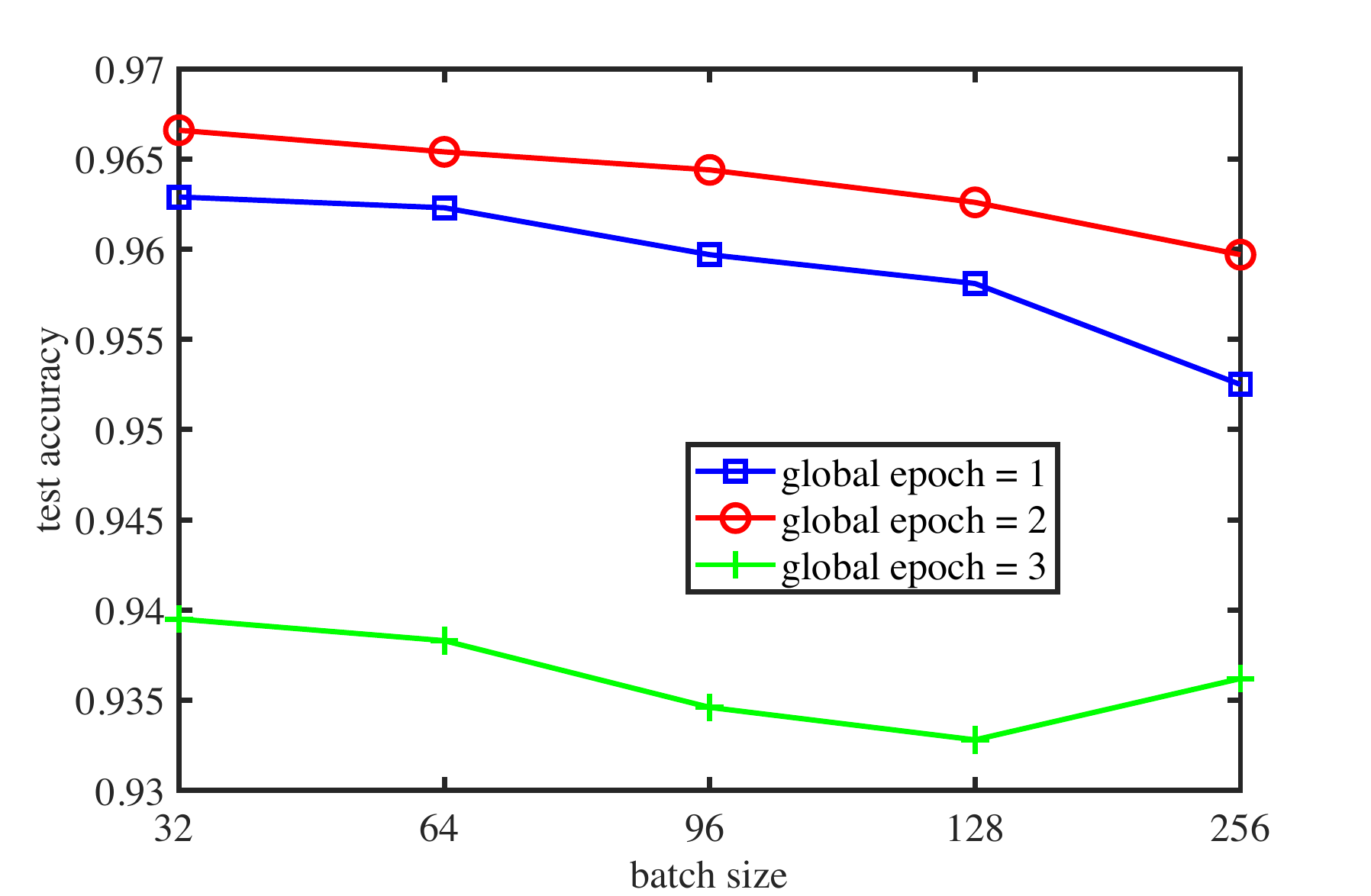}\vspace{-5pt}
\caption{%Impact of the global epoch and batch size on the test accuracy of the federated learning model under $\epsilon$-differential privacy for $\epsilon$ and local epochs being $2$ and $40$ respectively. 
Impact of the batch size on the test accuracy under different global epochs using our normalization technique ($\epsilon = 2$). \vspace{-12pt}}
\label{fig:FL-bz}
\end{figure}

Figure~\ref{fig:FL-dp-no} compares the test accuracies between federated learning (FL) without differential privacy (DP) and different DP-aware FL algorithms, including DP-aware FL using our normalization technique, and DP-aware FL using Jiang~\emph{et~al.}'s batch normalization~\cite{jiangdifferentially}.
Figure~\ref{fig:FL-dp-no} shows the superiority of DP-aware FL using our normalization technique over DP-aware FL using Jiang~\emph{et~al.}'s batch normalization~\cite{jiangdifferentially}. Thus, we confirm that our normalization technique is useful when we add Laplace noise to features, because we relax constraints of normalization compared with batch normalization as stated in Section~\ref{subsec:customers}. A feature goes through batch normalization often results in a smaller magnitude than that goes through  our normalization technique, so the value of feature is easily overwhelmed by the noise when using batch normalization. For each DP-aware FL, we also observe that the test accuracy gets closer to the test accuracy of FL without DP as the privacy parameter $\epsilon$ increases, because a larger privacy parameter $\epsilon$ means less privacy protection which equals that less noise is used. Thus, we conclude that our normalization technique outperforms the batch normalization under $\epsilon$-differential privacy when training the FL model.

% However, Figure~\ref{fig:FL-no-dp} shows that %if there is not any differential privacy noise
% when no differential privacy noise is added, batch normalization outperforms our normalization technique. Moreover, as the batch size increases, the test accuracy will decrease. Therefore, we conclude that our normalization technique is more practical since we ensure a higher accuracy while preserving privacy.%outperforms batch normalization when there is differential privacy noise, while the batch normalization %ensures 
% yields a better test accuracy than our normalization technique if there is not any noise.  

\begin{figure}[!t]
\centering
\includegraphics[width=\linewidth]{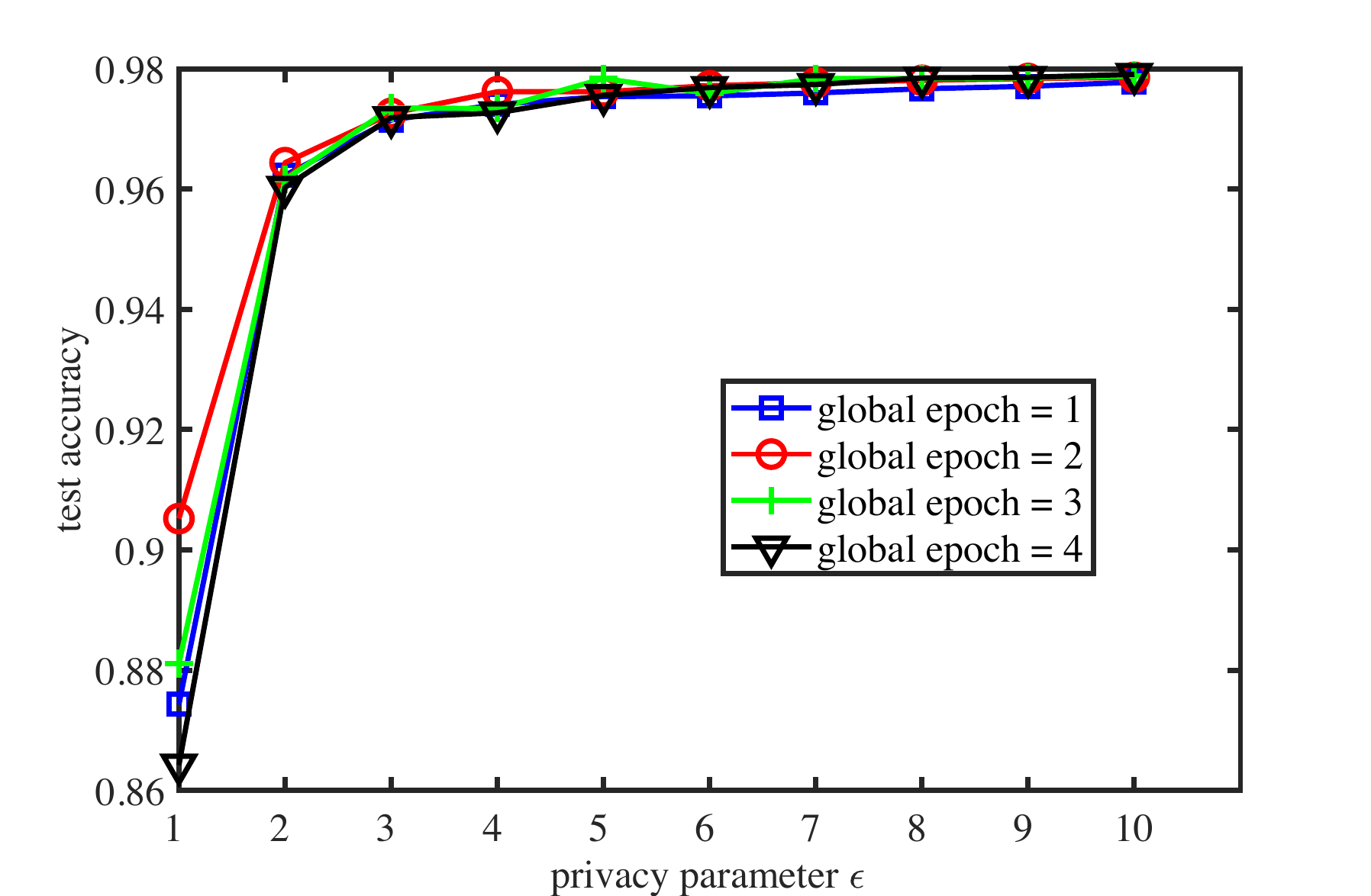}\vspace{-5pt}
\caption{%Impact of  differential privacy (DP) parameter $\epsilon$ on the test accuracy when the number of global epochs $= 1, 2, 3~\textup{and}~4$.
Impact of DP parameter $\epsilon$ on the test accuracy using our normalization technique under various global epochs.\vspace{-2pt}}
\label{fig:FL-dp}
\end{figure}

\begin{figure}[!t]
\centering
\includegraphics[width=\linewidth]{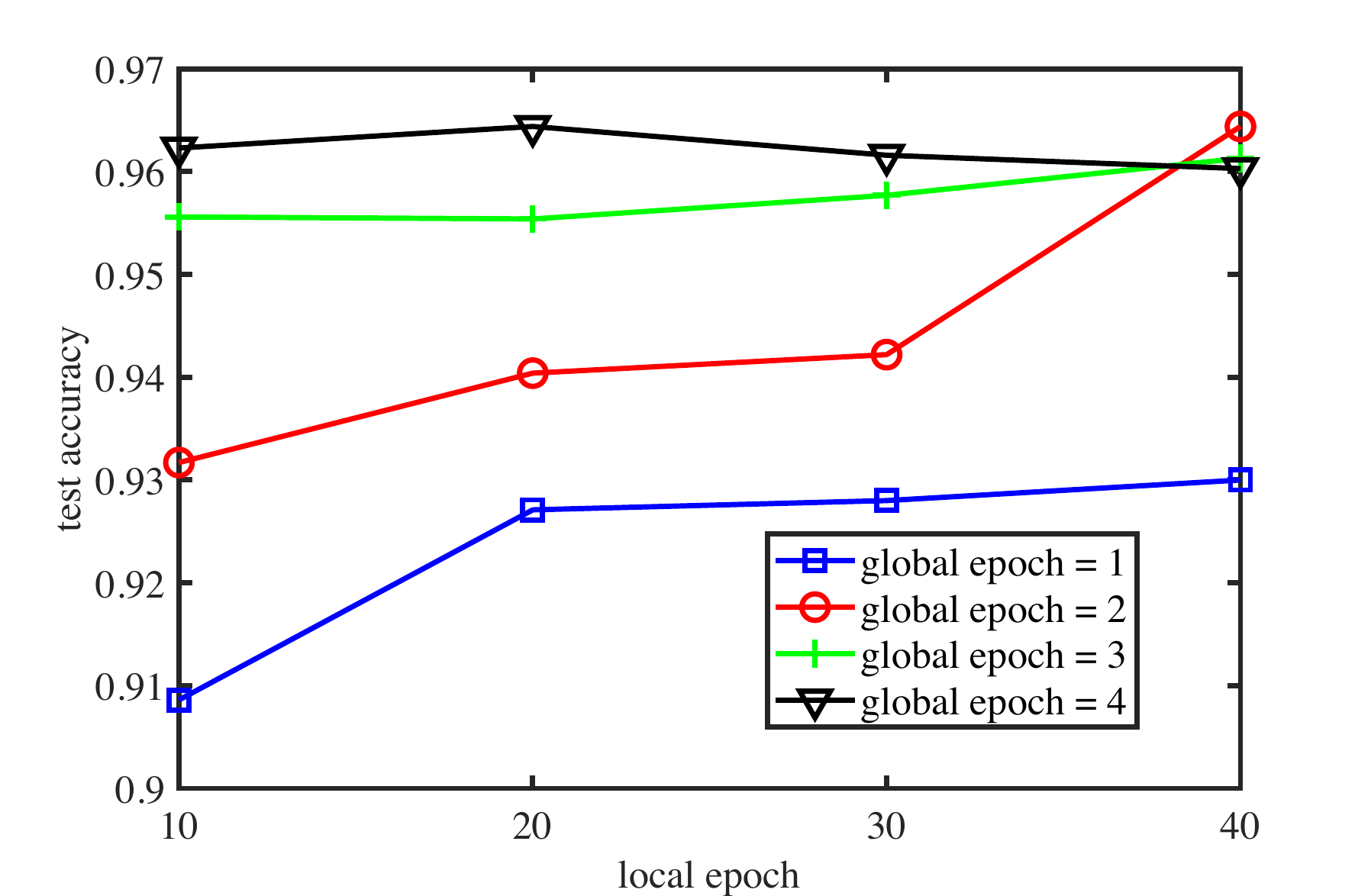}\vspace{-5pt}
\caption{Impact of the number of local epochs on the test accuracy using our normalization technique  under various global  epochs when $\epsilon = 2$.\vspace{-2pt}}
\label{fig:FL-epoch}
\end{figure}

Figure~\ref{fig:FL-bz-256} presents that the test accuracy of FL model decreases as the batch size increases  when the number of global epoch is $1$ and DP parameter $\epsilon = 2$. This is because we add Laplace noise to features, the added noise %added to features 
will increase as the batch size $N$ increases, which results in a worse test accuracy. Moreover, due to the three-sigma rule  in Gaussian distribution, most feature values normalized with batch normalization lie in $[-3\sigma, 3\sigma]$. But feature values normalized using our normalization technique lie in $[-\sqrt{N-1}, \sqrt{N-1}]$. However, Figure~\ref{fig:FL-no-dp} shows that if no differential privacy noise is added, the test accuracy with the batch normalization outperforms that using our normalization technique. Moreover, as the batch size increases, the test accuracy will decrease. Therefore, we conclude that our normalization technique works better with FL under DP protection.
%outperforms batch normalization when there is differential privacy noise, while the batch normalization %ensures 
% yields a better test accuracy than our normalization technique if there is not any noise. 
%Thus, it is equivalent to say 
%It implies that features values %after 
%normalized with batch normalization are smaller than %they 
% the features values normalized with our normalization technique. %Hence, feature values after batch normalization are gotten perturbed more easily than they are normalized by our normalization technique as we state in Section~\ref{subsection-explanation}. Thus, as shown in the Figure~\ref{fig:FL-bz-256}, when using Jiang~\emph{et~al.}'s~\cite{jiangdifferentially} batch normalization, test accuracies are worse than using our normalization technique. 
Furthermore, Figure~\ref{fig:FL-bz} illustrates that the test accuracy is better when the number of global epoch $ =2$ than the number of global epoch $= 1~\textup{or}~3$ when $\epsilon = 2$ and the number of local epochs is $40$. As the number of global epochs increases, the test accuracy increases  if DP noise is not added. However, Laplace noise increases as the number of global epochs increases, which negatively affects the test accuracy. Thus, a trade-off between the number of global epochs and the amount of noise is required. In our case, when the privacy parameter $\epsilon=2$ and the number of local epochs is $40$, the optimal number of global epochs is $2$.  

Figure~\ref{fig:FL-dp} illustrates how the privacy parameter $\epsilon$ affects the test accuracy of FL model. In our experiment, we train FL with $4$ global epochs to validate the practicality of our designed approach. %verify that our designed approach is practical. 
The test accuracy increases as the privacy parameter $\epsilon$ increases. A larger $\epsilon$ means that less noise is added to features, so that the privacy protection is weaker. Typical $\epsilon$ values for experiments are between $0.1$ and $10$~\cite{sanchez2014improving}. Our experiment shows that we can achieve at least $90\%$ accuracy when the global epoch $=2$ and the privacy parameter $\epsilon >1$. % is larger than $1$. 
Before training, we initialize the model with random parameters, and the model with initial parameters will be used by all parties for their local training. After the first global epoch, we obtain a new model by averaging all parties' model parameters. Then, in the second global epoch, parties start training using the model from the first global epoch. Through our experiment, we can verify that our designed FL method is effective. However, when the number of global epochs increases to $3$ or $4$, the test accuracy may decrease. The test accuracy decreases because the noise increases as the number of global epochs increases.

Figure~\ref{fig:FL-epoch} shows that the test accuracy of FL model is affected by both the number of local epochs and the number of the global epochs. The number of local epochs reflects the cost of devices' computing resources locally. %Normally, if a device trains the model with more epochs, and then the locally trained model will obtain a higher accuracy. Thus, the number of global epochs will reduce. 
We add $\epsilon$-differential privacy noise during training, and the test accuracy may drop if there is too much noise added in each epoch. From Figure~\ref{fig:FL-epoch}, when the number of local epochs equals $20$ or $30$, it takes $4$ global epochs to achieve a similar accuracy. When the number of local epochs is $40$, it takes $2$ global epochs.
But the test accuracy will start to drop if the number of local epochs is $40$ and the number of global epoch is more than $2$. Hence, to obtain a high test accuracy, %it is better 
it necessities optimal values to %find the 
strike a good balance between the number of local epochs and global epochs for averaging locally uploaded models, which we leave as the future work.

\subsection{Performance evaluation on the mobile device and edge server}

\begin{figure}[!htb]
 \centering
\minipage{0.17\textwidth}
  \includegraphics[width=\linewidth]{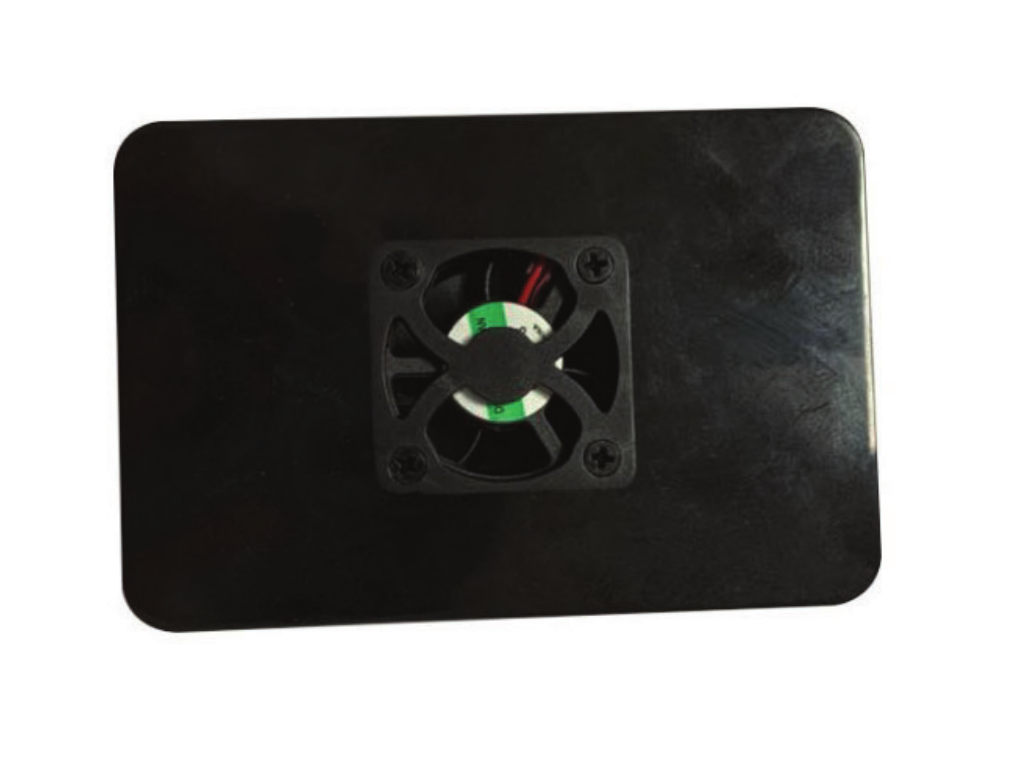}
\endminipage\hfill
 \centering
\minipage{0.23\textwidth}
  \includegraphics[width=\linewidth]{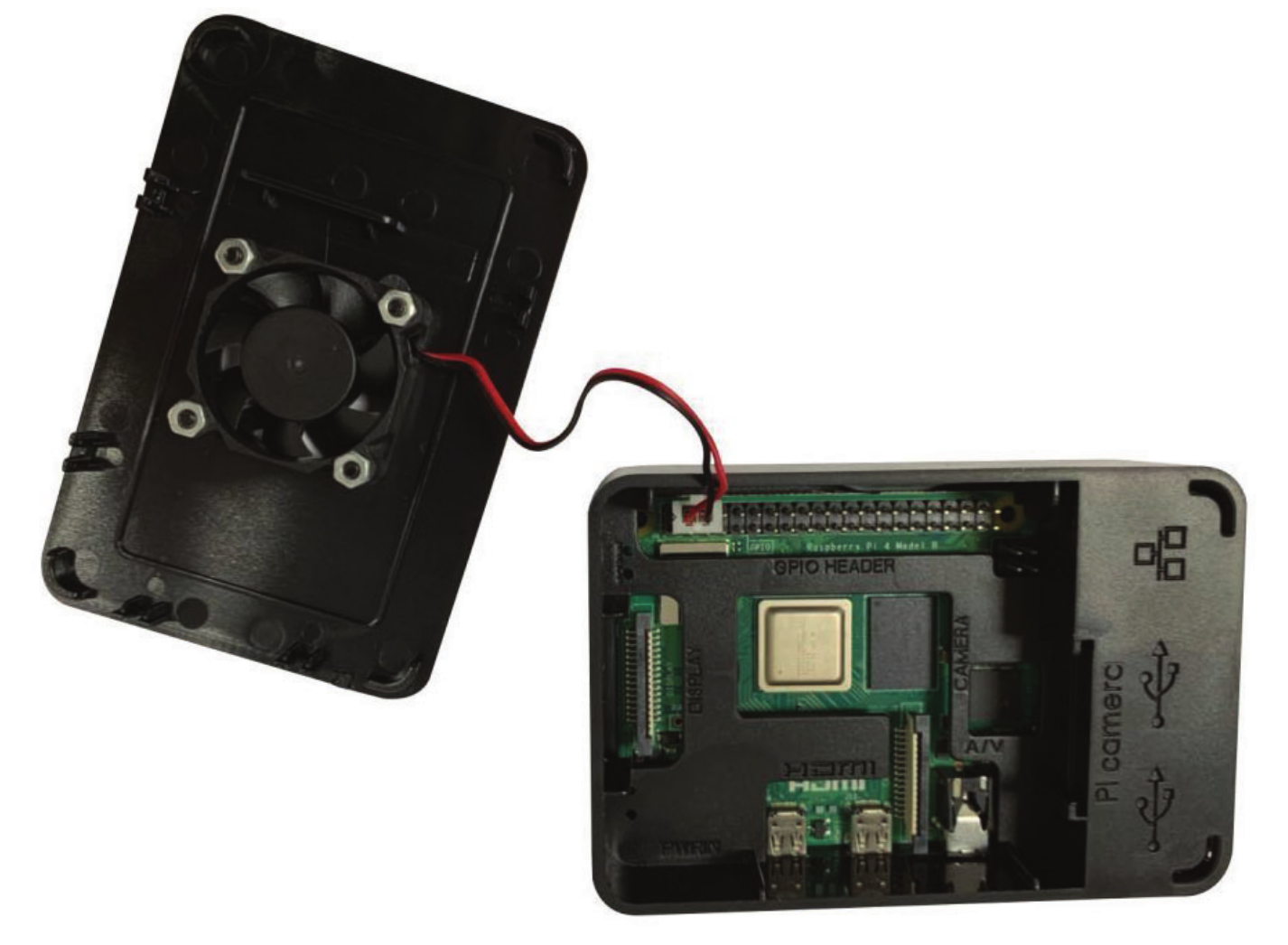}
\endminipage\hfill\hfill\hfill
 \centering
\minipage{0.15\textwidth}%
  \includegraphics[width=\linewidth]{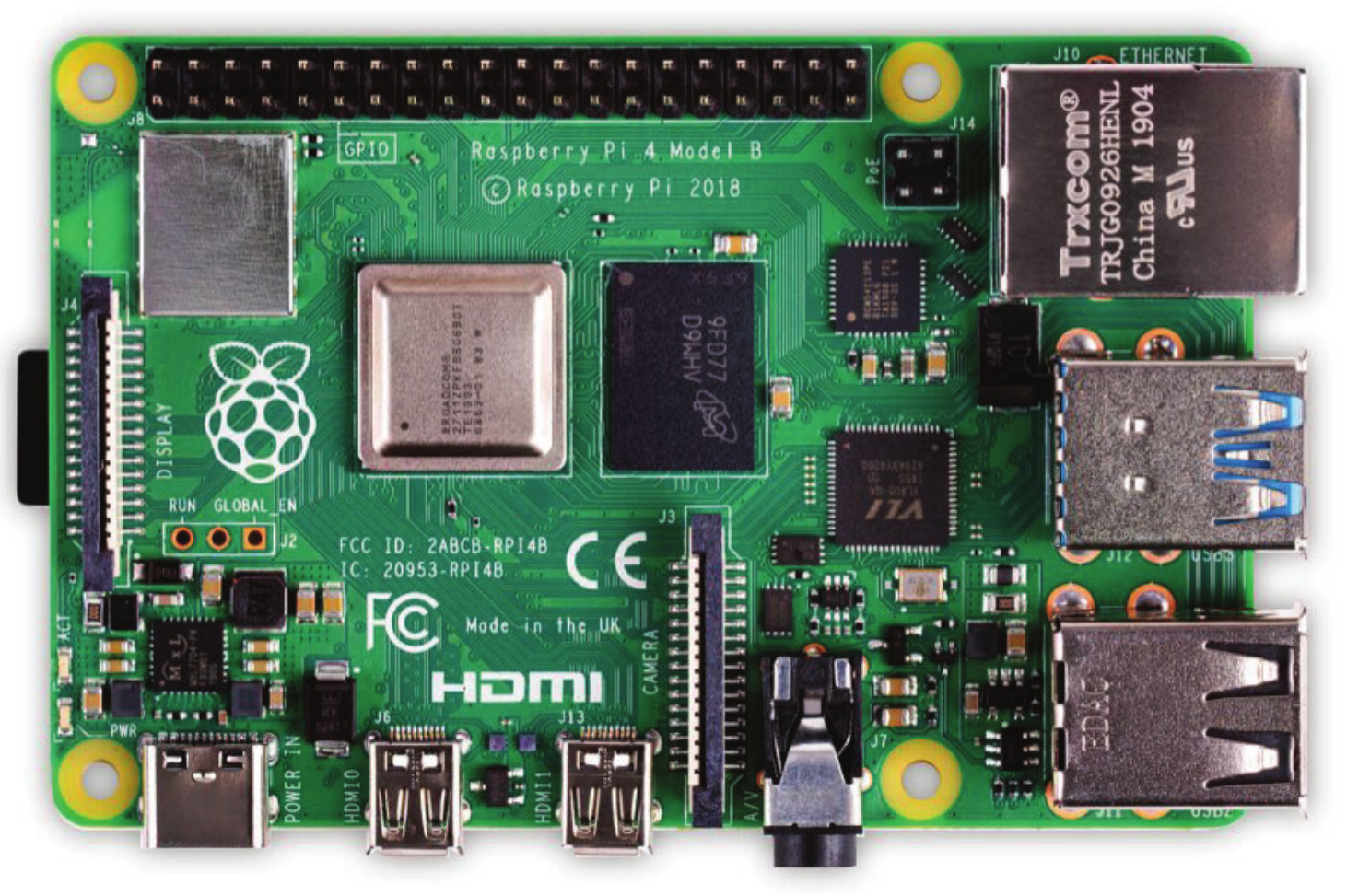}
\endminipage
  \caption{Raspberry Pi 4 Model B.}\label{fig:raspberry}
\end{figure}

Now, we evaluate the feasibility and efficacy of training on the mobile device. A Raspberry Pi $4$ Model B tiny computer in Figure~\ref{fig:raspberry} is used to simulate the mobile device. Key specifications of the Raspberry Pi $4$ Model B are listed in Table~\ref{table:raspberry}. We use a laptop to emulate the edge server, which is equipped with four 2.3 GHz Intel Core i5 processors, $8$ GB of RAM, and  MacOS 10.14.4 system.

\begin{table}[!htb]
\centering
\caption{Raspberry Pi 4 Model B Specifications~\cite{raspberrypi}.}\label{table:raspberry}
\begin{tabular}{|l|}
\hline
\begin{tabular}[c]{@{}l@{}}Broadcom BCM2711, Quad core Cortex-A72 (ARM v8) 64-bit SoC \\ @1.5GHz\end{tabular} \\ \hline
4GB LPDDR4-3200 SDRAM                                                                                          \\ \hline
\end{tabular}
\end{table}

In our experiment, we  distribute the MNIST dataset [52] with 60,000 images to ten participants equally so that each participant (i.e., each device) has 6,000 images. Then, we run the same training process on both the mobile device and the edge server. It takes about $144$ seconds to train the model with 6,000 images on the Raspberry Pi $4$ (i.e., the mobile device) for each epoch, and it uses about $9$ seconds to train the model on the laptop (i.e., the edge server). For default forty epochs, the mobile device and the edge server use about $96$ minutes and $6$ minutes, respectively. A client is supposed to participate in the federated learning when the smartphone is idle, such as charging, screen off, and connected to an unmetered network, for example, WiFi~\cite{mcmahan2017communication, bonawitz2019towards}. Thus, we confirm that it is feasible to utilize mobile devices in the federated learning. Besides, an edge server will significantly improve the speed of training because it trains much faster.

% \begin{figure}[!htb]
% \centering
% \includegraphics[width=\linewidth]{images/device.pdf}
% \caption{Comparison of training time on the mobile device and edge server.\vspace{-2pt}}
% \label{fig:device}
% \end{figure}

In addition to the training time, the delay of our proposed approach, which depends on the transmission rate, is small because smartphones often use wideband network connections (e.g., 4G and WiFi). The average size of locally trained models is $617.8$KB in our experiment.  Assume the upload bandwidth is 1MB/s, so the communication cost is $0.6178$ second. The communication cost is little compared with wasted training time on the mobile device.

\subsection{Evaluation on  the incentive mechanism}

In  this  section, we evaluate the impacts of incentive  mechanism on customers' reward and reputation. The assumption and parameters in the experiments are as follows.  Assume that the maximum values of both reputation and reward are $100$  (i.e., $\gamma^{Max} = 100$). Every customer has a reputation of $5$ (i.e., $h = 5$) at the beginning.  We set the reward for each accepted update equal to owners' reputation in each global epoch. The experiments compare reward and reputation that customer can achieve in four cases (i.e., no incentive mechanism, honest customer, malicious customer performs poisoning attack at global epoch = 1, and malicious customer performs poisoning attack at global epoch = 4). If there is no incentive mechanism, the customer gets a fixed reward of $5$ in every global epoch.

\begin{figure}[!h]
    \centering
    \includegraphics[width=\linewidth]{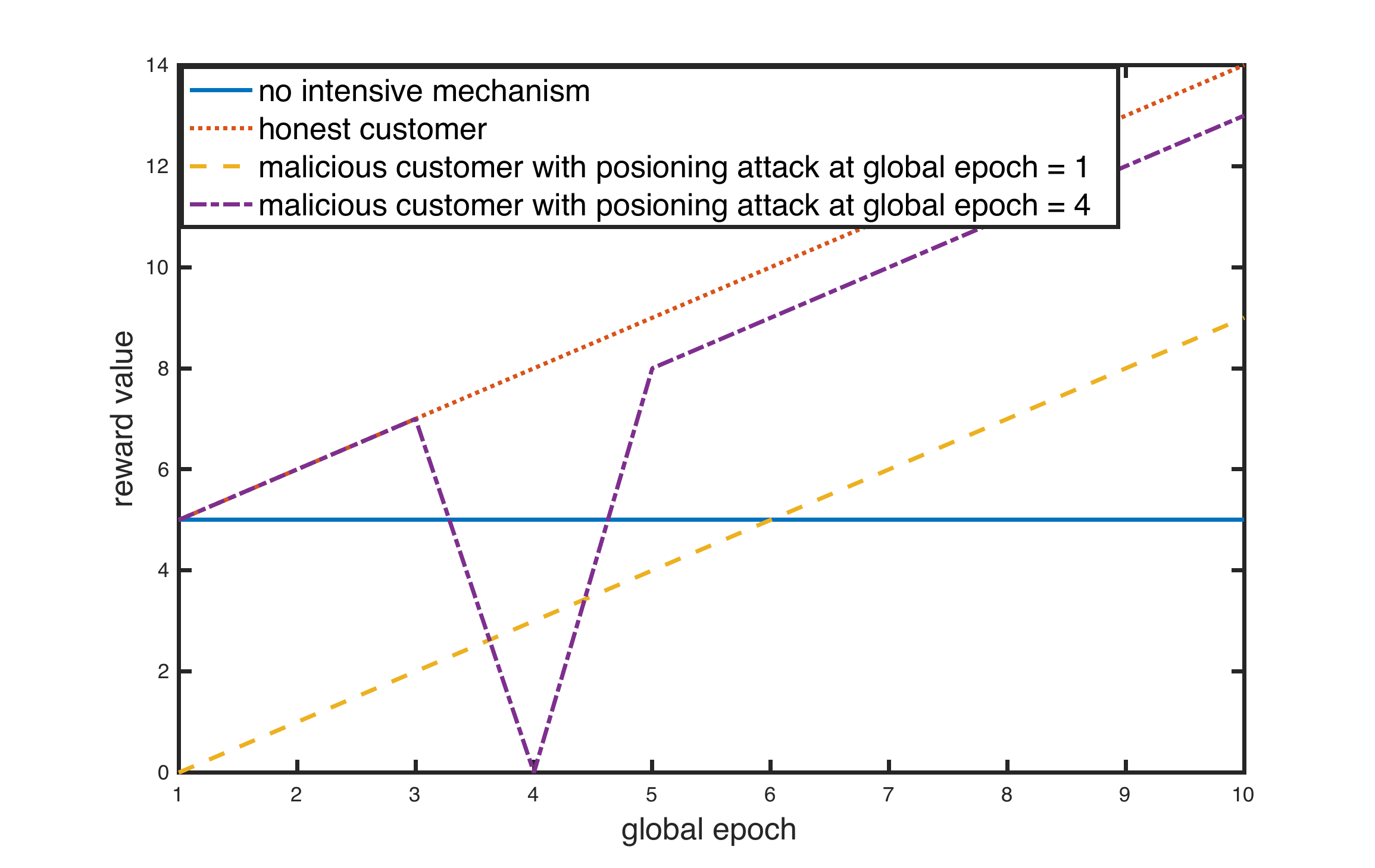}
    \caption{Reward comparison.}
    \label{fig:reward}
\end{figure}

As shown in Figure~\ref{fig:reward}, when there is no incentive mechanism, the reward value is the same in each global epoch regardless of poisoning updates. However, with the incentive mechanism, the honest customer, whose updates are accepted, will gain more rewards as the number of global epochs increases. If a customer's update is considered as poisoning (i.e., the value of $s$ in  Eq.~(\ref{eq:distance}) is significantly larger than  others), her update will not be accepted, that is, her reward is $0$. Besides, the behaviour of the poisoning attack affects the value of reputation, which results in a decrease of the reputation. If the poisoning attack is performed when the value of the reputation is equal to the $h$, the customer's  reputation will be clear, which will result in small rewards afterwards. However, if the malicious behaviour happens when the  value of reputation is higher than  $h$, the reputation drops by $1$, so does the reward in the subsequent global epoch.

\begin{figure}[!h]
    \centering
    \includegraphics[width=\linewidth]{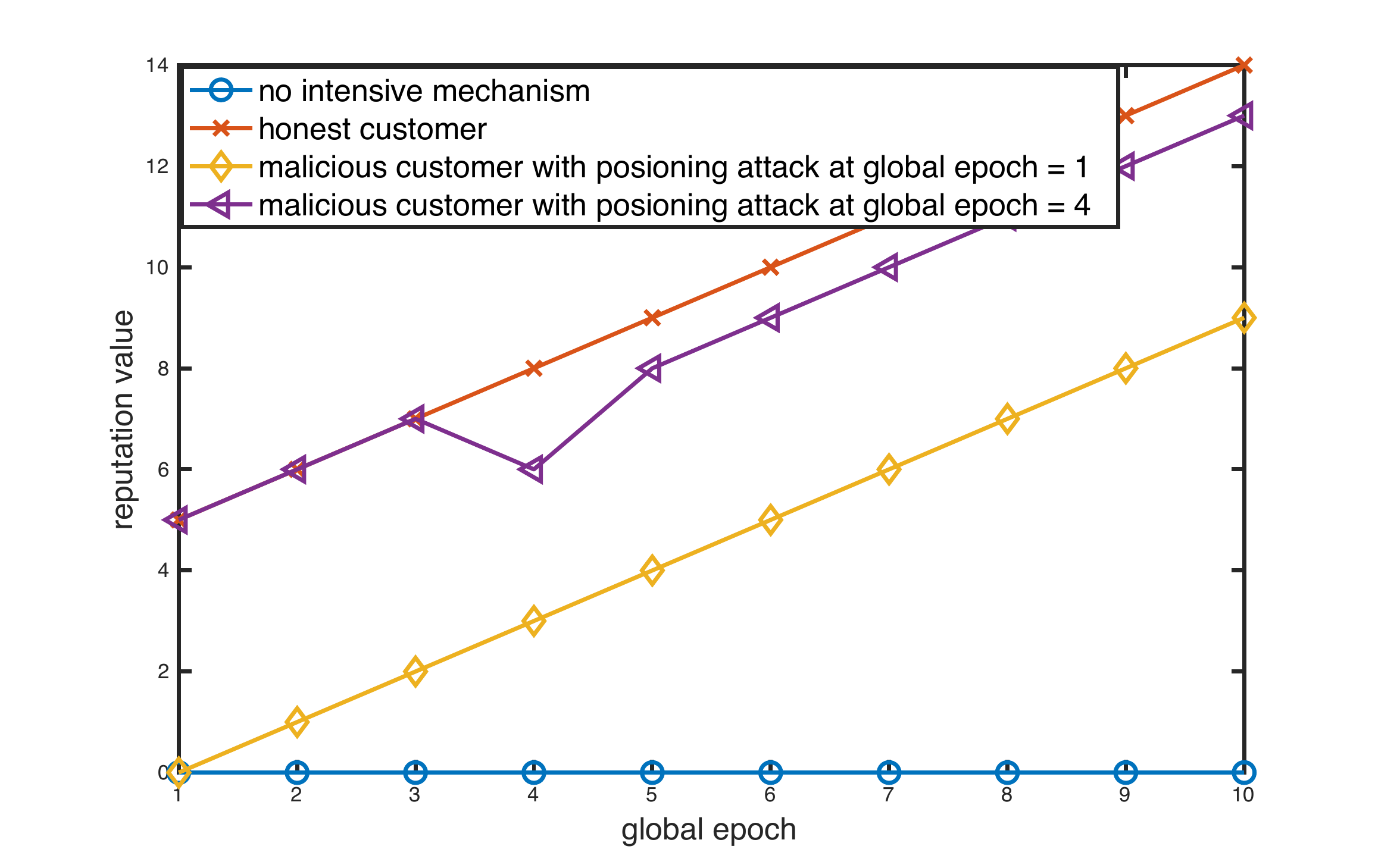}
    \caption{Reputation comparison.}
    \label{fig:reputation}
\end{figure}

Figure~\ref{fig:reputation} shows the impact of the incentive mechanism on the reputation. Without the incentive mechanism, customers' reputation will be  $0$. If  a customer is honest  and uploads the correct update  in every global epoch, her reputation increases  as the number of global epoch increases. However, if a customer uploads a malicious update when her reputation value equals to the  $h$ (i.e., $5$), her reputation will drop to $0$. However, if her reputation is  not $5$, her reputation drops by $1$ when caught performing poisoning attack. 

Thus, our incentive mechanism can encourage honest customers to contribute their useful updates while preventing malicious customers from attempting to perform the poisoning attack.

\section{Discussion}~\label{sec-discussion}
To attract more customers to contribute to training the global model, our designed system should guarantee that customers' confidential information will not leak. There are studies discussing potential risks of information leakage in FL~\cite{hitaj2017deep, yin2019dreaming} in which attackers may infer customers' private data from gradients. To prevent this scenario, we leverage differential privacy %in 
to perturb features before classification in the fully connected layers. Thus, gradients are also protected by differential privacy. Hitaj~\emph{et~al.}~\cite{hitaj2017deep} demonstrated that a curious server could obtain the confidential information using the generative adversarial network if gradients were protected by a large privacy budget in the collaborative learning. But their experiments confirmed that GAN-based approaches might not work well when the selected privacy parameter was smaller than $10$, which is the upper bound of privacy parameter in our experiment, %. Hence, the privacy parameter assigned to the each gradient is relatively small, but 
and we can achieve the accuracy of $97\%$. Therefore, our designed approach guarantees the accuracy and protects the privacy of local models as well as data. In addition, Yin~\emph{et~al.}~\cite{yin2019dreaming} introduced a DeepInversion method which could invert a trained neural network to a synthesized class-conditional input images starting from the random noise. However, they leveraged the information stored in the batch normalization layer. In our trained model, we add Laplacian noise during training in the batch normalization layer, and then attackers are unable to obtain the true information stored there. Thus, their approach is ineffective against our trained model.

% We discuss advantages and disadvantages of our framework in this section.
% \subsection{Privacy and Security}
% Our system leverages differential privacy technique to protect the privacy of the extracted features.
% Thus, the system keeps the participating customers' data confidential. 
% Furthermore, the trained model is encrypted and signed by the sender to prevent the attackers and imposters from stealing the model or deriving original data through reverse-engineering.

% \subsection{Delay in Crowdsourcing}
% Assume there is a large number of customers, and the system highly depends on customers' training results to obtain the predictive model in one global epoch. Unlike other crowdsourcing jobs, manufacturers in our system prefers customers to follow their lifestyle instead of rushing to finish the job to obtain the real status. As a result, customers who seldom use devices may postpone the overall crowdsourcing progress. This problem can be mitigated by using the incentive mechanisms.  Yu~\emph{et~al.}~\cite{yu2020fairness} designed a queue to store customers who submitted their models in order. Thus, customers who submit their locally trained models early will be rewarded to encourage people to submit their updates earlier.

\section{Conclusion and Future Work}\label{sec-conclusion}
In this paper, we present a design of blockchain-based crowdsourcing FL system for IoT devices manufacturers to learn customers better. We use multiple \mbox{state-of-the-art}  technologies to construct the system, including the mobile edge computing server, blockchain, distributed storage, and federated learning. Besides, our system enforces differential privacy to protect the privacy of customers' data. To improve the accuracy of  FL model, we design a new normalization technique which is proved to outperform the batch normalization if features' privacy is protected by differential privacy.  By designing a proper incentive mechanism for the crowdsourcing task, customers are
more likely to participate in the crowdsourcing tasks. The blockchain will audit all customers' updates during the federated training, so that the system can hold the model updates accountable to prevent malicious customers or manufacturers.

In the future, we aim to conduct more experiments and test our system with real-world home appliance datasets. Moreover, we will strive to find the deterministically optimal balance between local epochs and global epochs to obtain a better test accuracy.

\bibliographystyle{IEEEtran}
\bibliography{reference}

\begin{thebibliography}{10}
\providecommand{\url}[1]{#1}
\csname url@samestyle\endcsname
\providecommand{\newblock}{\relax}
\providecommand{\bibinfo}[2]{#2}
\providecommand{\BIBentrySTDinterwordspacing}{\spaceskip=0pt\relax}
\providecommand{\BIBentryALTinterwordstretchfactor}{4}
\providecommand{\BIBentryALTinterwordspacing}{\spaceskip=\fontdimen2\font plus
\BIBentryALTinterwordstretchfactor\fontdimen3\font minus
  \fontdimen4\font\relax}
\providecommand{\BIBforeignlanguage}[2]{{%
\expandafter\ifx\csname l@#1\endcsname\relax
\typeout{** WARNING: IEEEtran.bst: No hyphenation pattern has been}%
\typeout{** loaded for the language `#1'. Using the pattern for}%
\typeout{** the default language instead.}%
\else
\language=\csname l@#1\endcsname
\fi
#2}}
\providecommand{\BIBdecl}{\relax}
\BIBdecl

\bibitem{smarthomesta}
\BIBentryALTinterwordspacing
S.~R. Department, ``Smart home - {S}tatistics \& {F}acts,'' 2020. [Online].
  Available: \url{https://www.statista.com/topics/2430/smart-homes/}
\BIBentrySTDinterwordspacing

\bibitem{wang2018adaptive}
S.~Wang, T.~Tuor, T.~Salonidis, K.~K. Leung, C.~Makaya, T.~He, and K.~Chan,
  ``When edge meets learning: {Adaptive} control for resource-constrained
  distributed machine learning,'' in \emph{IEEE Conference on Computer
  Communications (INFOCOM)}, 2018, pp. 63--71.

\bibitem{melis2018inference}
L.~Melis, C.~Song, E.~De~Cristofaro, and V.~Shmatikov, ``Exploiting unintended
  feature leakage in collaborative learning,'' in \emph{IEEE Symposium on
  Security and Privacy (S\&P)}, 2019.

\bibitem{hitaj2017deep}
B.~Hitaj, G.~Ateniese, and F.~Perez-Cruz, ``Deep models under the {GAN}:
  information leakage from collaborative deep learning,'' in \emph{Proceedings
  of the 2017 ACM SIGSAC Conference on Computer and Communications Security},
  2017, pp. 603--618.

\bibitem{fung2018mitigating}
C.~Fung, C.~J. Yoon, and I.~Beschastnikh, ``Mitigating sybils in federated
  learning poisoning,'' \emph{arXiv preprint arXiv:1808.04866}, 2018.

\bibitem{zhang2020mdldroid}
Y.~Zhang, T.~Gu, and X.~Zhang, ``Mdldroid: a chainsgd-reduce approach to mobile
  deep learning for personal mobile sensing,'' in \emph{2020 19th ACM/IEEE
  International Conference on Information Processing in Sensor Networks
  (IPSN)}.\hskip 1em plus 0.5em minus 0.4em\relax IEEE, 2020, pp. 73--84.

\bibitem{appledifferentialprivacy}
\BIBentryALTinterwordspacing
K.~Hao, ``How {A}pple personalizes {S}iri without hoovering up your data,''
  2019. [Online]. Available:
  \url{https://www.technologyreview.com/2019/12/11/131629/apple-ai-personalizes-siri-federated-learning/}
\BIBentrySTDinterwordspacing

\bibitem{benet2014ipfs}
J.~Benet, ``{IPFS}-content addressed, versioned, {P2P} file system,''
  \emph{arXiv preprint arXiv:1407.3561}, 2014.

\bibitem{dwork2006calibrating}
C.~Dwork, F.~McSherry, K.~Nissim, and A.~Smith, ``Calibrating noise to
  sensitivity in private data analysis,'' in \emph{Theory of Cryptography
  Conference (TCC)}, 2006, pp. 265--284.

\bibitem{dwork2006our}
C.~Dwork, K.~Kenthapadi, F.~McSherry, I.~Mironov, and M.~Naor, ``Our data,
  ourselves: Privacy via distributed noise generation,'' in \emph{International
  Conference on the Theory and Applications of Cryptographic Techniques
  (EUROCRYPT)}, 2006, pp. 486--503.

\bibitem{tang2017privacy}
J.~Tang, A.~Korolova, X.~Bai, X.~Wang, and X.~Wang, ``Privacy loss in {A}pple's
  implementation of differential privacy on {macOS} 10.12,'' \emph{arXiv
  preprint arXiv:1709.02753}, 2017.

\bibitem{erlingsson2014rappor}
{\'U}.~Erlingsson, V.~Pihur, and A.~Korolova, ``{RAPPOR}: Randomized
  aggregatable privacy-preserving ordinal response,'' in \emph{ACM Conference
  on Computer and Communications Security (CCS)}, 2014, pp. 1054--1067.

\bibitem{mcmahan2017communication}
B.~McMahan, E.~Moore, D.~Ramage, S.~Hampson, and B.~A. y~Arcas,
  ``Communication-efficient learning of deep networks from decentralized
  data,'' in \emph{Artificial Intelligence and Statistics}, 2017, pp.
  1273--1282.

\bibitem{konevcny2016federated}
J.~Kone{\v{c}}n{\`y}, H.~B. McMahan, F.~X. Yu, P.~Richt{\'a}rik, A.~T. Suresh,
  and D.~Bacon, ``Federated learning: Strategies for improving communication
  efficiency,'' in \emph{NIPS Workshop on Private Multi-Party Machine
  Learning}, 2016.

\bibitem{qu2019proof}
X.~Qu, S.~Wang, Q.~Hu, and X.~Cheng, ``Proof of federated learning: A novel
  energy-recycling consensus algorithm,'' \emph{arXiv preprint
  arXiv:1912.11745}, 2019.

\bibitem{lu2019blockchain}
Y.~Lu, X.~Huang, Y.~Dai, S.~Maharjan, and Y.~Zhang, ``Blockchain and federated
  learning for privacy-preserved data sharing in industrial {IoT},'' \emph{IEEE
  Transactions on Industrial Informatics}, 2019.

\bibitem{ramanan2019baffle}
P.~Ramanan, K.~Nakayama, and R.~Sharma, ``Baffle: Blockchain based aggregator
  free federated learning,'' \emph{arXiv preprint arXiv:1909.07452}, 2019.

\bibitem{kim2019blockchained}
H.~Kim, J.~Park, M.~Bennis, and S.-L. Kim, ``Blockchained on-device federated
  learning,'' \emph{IEEE Communications Letters}, 2019.

\bibitem{yin2020fdc}
B.~Yin, H.~Yin, Y.~Wu, and Z.~Jiang, ``{FDC}: A secure federated deep learning
  mechanism for data collaborations in the {I}nternet of {T}hings,'' \emph{IEEE
  {I}nternet of {T}hings Journal}, 2020.

\bibitem{awan2019poster}
S.~Awan, F.~Li, B.~Luo, and M.~Liu, ``Poster: A reliable and accountable
  privacy-preserving federated learning framework using the blockchain,'' in
  \emph{Proceedings of the 2019 ACM SIGSAC Conference on Computer and
  Communications Security}, 2019, pp. 2561--2563.

\bibitem{weng2019deepchain}
J.~Weng, J.~Weng, J.~Zhang, M.~Li, Y.~Zhang, and W.~Luo, ``Deep{C}hain:
  Auditable and privacy-preserving deep learning with blockchain-based
  incentive,'' \emph{IEEE Transactions on Dependable and Secure Computing},
  2019.

\bibitem{lyu2019towards}
L.~Lyu, J.~Yu, K.~Nandakumar, Y.~Li, X.~Ma, and J.~Jin, ``Towards fair and
  decentralized privacy-preserving deep learning with blockchain,'' \emph{arXiv
  preprint arXiv:1906.01167}, 2019.

\bibitem{ZhaoWZZC19}
L.~Zhao, Q.~Wang, Q.~Zou, Y.~Zhang, and Y.~Chen, ``Privacy-preserving
  collaborative deep learning with unreliable participants,'' \emph{IEEE
  Transactions on Information Forensics and Security}, vol.~15, pp. 1486--1500,
  2019.

\bibitem{ZhaoHWJSLH20}
L.~Zhao, S.~Hu, Q.~Wang, J.~Jiang, C.~Shen, X.~Luo, and P.~Hu, ``Shielding
  collaborative learning: Mitigating poisoning attacks through client-side
  detection,'' \emph{{IEEE} Transactions on Dependable and Secure Computing},
  vol.~PP, no.~99, pp. 1--1, 10.1109/TDSC.2020.2\,986\,205, 2020.

\bibitem{li2018crowdbc}
M.~Li, J.~Weng, A.~Yang, W.~Lu, Y.~Zhang, L.~Hou, J.-N. Liu, Y.~Xiang, and
  R.~Deng, ``{CrowdBC}: A blockchain-based decentralized framework for
  crowdsourcing,'' \emph{IEEE Transactions on Parallel and Distributed
  Systems}, 2018.

\bibitem{yang2019federated}
Q.~Yang, Y.~Liu, T.~Chen, and Y.~Tong, ``Federated machine learning: Concept
  and applications,'' \emph{ACM Transactions on Intelligent Systems and
  Technology (TIST)}, vol.~10, no.~2, pp. 1--19, 2019.

\bibitem{lim2019federated}
W.~Y.~B. Lim, N.~C. Luong, D.~T. Hoang, Y.~Jiao, Y.-C. Liang, Q.~Yang,
  D.~Niyato, and C.~Miao, ``Federated learning in mobile edge networks: A
  comprehensive survey,'' \emph{IEEE Communications Surveys \& Tutorials},
  2020.

\bibitem{li2019federated}
T.~Li, A.~K. Sahu, A.~Talwalkar, and V.~Smith, ``Federated learning:
  Challenges, methods, and future directions,'' \emph{IEEE Signal Processing
  Magazine}, vol.~37, no.~3, pp. 50--60, 2020.

\bibitem{nishio2019client}
T.~Nishio and R.~Yonetani, ``Client selection for federated learning with
  heterogeneous resources in mobile edge,'' in \emph{ICC 2019-2019 IEEE
  International Conference on Communications (ICC)}.\hskip 1em plus 0.5em minus
  0.4em\relax IEEE, 2019, pp. 1--7.

\bibitem{lyu2020threats}
L.~Lyu, H.~Yu, and Q.~Yang, ``Threats to federated learning: A survey,''
  \emph{arXiv preprint arXiv:2003.02133}, 2020.

\bibitem{liu2019enhancing}
Z.~Liu, T.~Li, V.~Smith, and V.~Sekar, ``Enhancing the privacy of federated
  learning with sketching,'' \emph{arXiv preprint arXiv:1911.01812}, 2019.

\bibitem{hao2019efficient}
M.~Hao, H.~Li, X.~Luo, G.~Xu, H.~Yang, and S.~Liu, ``Efficient and
  privacy-enhanced federated learning for industrial artificial intelligence,''
  \emph{IEEE Transactions on Industrial Informatics}, 2019.

\bibitem{dolui2019poster}
K.~Dolui, I.~C. Gyllensten, D.~Lowet, S.~Michiels, H.~Hallez, and D.~Hughes,
  ``Poster: Towards privacy-preserving mobile applications with federated
  learning--the case of matrix factorization,'' in \emph{The 17th Annual
  International Conference on Mobile Systems, Applications, and Services, Date:
  2019/06/17-2019/06/21, Location: Seoul, Korea}, 2019.

\bibitem{nasr2019comprehensive}
M.~Nasr, R.~Shokri, and A.~Houmansadr, ``Comprehensive privacy analysis of deep
  learning: Passive and active white-box inference attacks against centralized
  and federated learning,'' in \emph{2019 IEEE Symposium on Security and
  Privacy (SP)}.\hskip 1em plus 0.5em minus 0.4em\relax IEEE, 2019, pp.
  739--753.

\bibitem{wang2019beyond}
Z.~Wang, M.~Song, Z.~Zhang, Y.~Song, Q.~Wang, and H.~Qi, ``Beyond inferring
  class representatives: User-level privacy leakage from federated learning,''
  in \emph{IEEE INFOCOM 2019-IEEE Conference on Computer Communications}.\hskip
  1em plus 0.5em minus 0.4em\relax IEEE, 2019, pp. 2512--2520.

\bibitem{wu2020crowdprivacy}
F.-J. Wu and T.~Luo, ``Crowd{P}rivacy: Publish more useful data with less
  privacy exposure in crowdsourced location-based services,'' \emph{ACM
  Transactions on Privacy and Security (TOPS)}, vol.~23, no.~1, pp. 1--25,
  2020.

\bibitem{liang2020enabling}
T.~Liang, ``Enabling privacy preservation and decentralization for
  attribute-based task assignment in crowdsourcing,'' \emph{Journal of Computer
  and Communications}, vol.~8, no.~4, pp. 81--100, 2020.

\bibitem{he2020privbus}
Y.~He, J.~Ni, B.~Niu, F.~Li, and X.~S. Shen, ``Privbus: A privacy-enhanced
  crowdsourced bus service via fog computing,'' \emph{Journal of Parallel and
  Distributed Computing}, vol. 135, pp. 156--168, 2020.

\bibitem{zhang2020fog}
J.~Zhang, Q.~Zhang, and S.~Ji, ``A fog-assisted privacy-preserving task
  allocation in crowdsourcing,'' \emph{IEEE {I}nternet of {T}hings Journal},
  2020.

\bibitem{zhao2020p3}
P.~Zhao, H.~Huang, X.~Zhao, and D.~Huang, ``P$^3$: Privacy-preserving scheme
  against poisoning attacks in mobile-edge computing,'' \emph{IEEE Transactions
  on Computational Social Systems}, 2020.

\bibitem{xu2019blockchain}
J.~Xu, S.~Wang, B.~K. Bhargava, and F.~Yang, ``A blockchain-enabled trustless
  crowd-intelligence ecosystem on mobile edge computing,'' \emph{IEEE
  Transactions on Industrial Informatics}, vol.~15, no.~6, pp. 3538--3547,
  2019.

\bibitem{shi2016edge}
W.~Shi, J.~Cao, Q.~Zhang, Y.~Li, and L.~Xu, ``Edge computing: Vision and
  challenges,'' \emph{IEEE {I}nternet of {T}hings journal}, vol.~3, no.~5, pp.
  637--646, 2016.

\bibitem{shi2016promise}
W.~Shi and S.~Dustdar, ``The promise of edge computing,'' \emph{Computer},
  vol.~49, no.~5, pp. 78--81, 2016.

\bibitem{lyu2019fog}
L.~Lyu, J.~C. Bezdek, X.~He, and J.~Jin, ``Fog-embedded deep learning for the
  {I}nternet of {T}hings,'' \emph{IEEE Transactions on Industrial Informatics},
  2019.

\bibitem{jiangdifferentially}
L.~Jiang, X.~Lou, R.~Tan, and J.~Zhao, ``Differentially private collaborative
  learning for the {{IoT}} edge,'' in \emph{International Workshop on Crowd
  Intelligence for Smart Cities: Technology and Applications (CISC)}, 2018.

\bibitem{wang2019adaptive}
S.~Wang, T.~Tuor, T.~Salonidis, K.~K. Leung, C.~Makaya, T.~He, and K.~Chan,
  ``Adaptive federated learning in resource constrained edge computing
  systems,'' \emph{IEEE Journal on Selected Areas in Communications}, vol.~37,
  no.~6, pp. 1205--1221, 2019.

\bibitem{wang2019edge}
X.~Wang, Y.~Han, C.~Wang, Q.~Zhao, X.~Chen, and M.~Chen, ``In-edge ai:
  Intelligentizing mobile edge computing, caching and communication by
  federated learning,'' \emph{IEEE Network}, vol.~33, no.~5, pp. 156--165,
  2019.

\bibitem{mao2018learning}
Y.~Mao, S.~Yi, Q.~Li, J.~Feng, F.~Xu, and S.~Zhong, ``Learning from
  differentially private neural activations with edge computing,'' in
  \emph{2018 IEEE/ACM Symposium on Edge Computing (SEC)}.\hskip 1em plus 0.5em
  minus 0.4em\relax IEEE, 2018, pp. 90--102.

\bibitem{wang2019survey}
W.~Wang, D.~T. Hoang, P.~Hu, Z.~Xiong, D.~Niyato, P.~Wang, Y.~Wen, and D.~I.
  Kim, ``A survey on consensus mechanisms and mining strategy management in
  blockchain networks,'' \emph{IEEE Access}, vol.~7, pp. 22\,328--22\,370,
  2019.

\bibitem{gilad2017algorand}
Y.~Gilad, R.~Hemo, S.~Micali, G.~Vlachos, and N.~Zeldovich, ``Algorand:
  {Scaling} byzantine agreements for cryptocurrencies,'' in \emph{ACM Symposium
  on Operating Systems Principles (SOSP)}, 2017, pp. 51--68.

\bibitem{blanchard2017machine}
P.~Blanchard, E.~M.~E. Mhamdi, R.~Guerraoui, and J.~Stainer, ``Machine learning
  with adversaries: Byzantine tolerant gradient descent,'' in \emph{Advances in
  Neural Information Processing Systems}, 2017, pp. 119--129.

\bibitem{shayan2018biscotti}
M.~Shayan, C.~Fung, C.~J. Yoon, and I.~Beschastnikh, ``Biscotti: A ledger for
  private and secure peer-to-peer machine learning,'' \emph{arXiv preprint
  arXiv:1811.09904}, 2018.

\bibitem{zhang2012reputation}
Y.~Zhang and M.~Van~der Schaar, ``Reputation-based incentive protocols in
  crowdsourcing applications,'' in \emph{2012 Proceedings IEEE INFOCOM}.\hskip
  1em plus 0.5em minus 0.4em\relax IEEE, 2012, pp. 2140--2148.

\bibitem{pukelsheim1994three}
F.~Pukelsheim, ``The three sigma rule,'' \emph{The American Statistician},
  vol.~48, no.~2, pp. 88--91, 1994.

\bibitem{yu2020fairness}
H.~Yu, Z.~Liu, Y.~Liu, T.~Chen, M.~Cong, X.~Weng, D.~Niyato, and Q.~Yang, ``A
  fairness-aware incentive scheme for federated learning,'' in
  \emph{Proceedings of the AAAI/ACM Conference on AI, Ethics, and Society},
  2020, pp. 393--399.

\bibitem{lecun2010mnist}
\BIBentryALTinterwordspacing
Y.~LeCun, C.~Cortes, and C.~Burges, ``{MNIST} handwritten digit database,''
  2010, accessed on March 1, 2019. [Online]. Available:
  \url{http://yann.lecun.com/exdb/mnist}
\BIBentrySTDinterwordspacing

\bibitem{xu2020lightweight}
D.~Xu, M.~Zheng, L.~Jiang, C.~Gu, R.~Tan, and P.~Cheng, ``Lightweight and
  unobtrusive data obfuscation at {IoT} edge for remote inference,'' \emph{IEEE
  {I}nternet of {T}hings Journal}, 2020.

\bibitem{zheng2019challenges}
M.~Zheng, D.~Xu, L.~Jiang, C.~Gu, R.~Tan, and P.~Cheng, ``Challenges of
  privacy-preserving machine learning in {IoT},'' in \emph{Proceedings of the
  First International Workshop on Challenges in Artificial Intelligence and
  Machine Learning for {I}nternet of {T}hings}, 2019, pp. 1--7.

\bibitem{mills2019communication}
J.~Mills, J.~Hu, and G.~Min, ``Communication-efficient federated learning for
  wireless edge intelligence in {IoT},'' \emph{IEEE {I}nternet of {T}hings
  Journal}, 2019.

\bibitem{liu2019fitcnn}
D.~Liu, C.~Yang, S.~Li, X.~Chen, J.~Ren, R.~Liu, M.~Duan, Y.~Tan, and L.~Liang,
  ``{FitCNN}: A cloud-assisted and low-cost framework for updating {CNN}s on
  {IoT} devices,'' \emph{Future Generation Computer Systems}, vol.~91, pp.
  277--289, 2019.

\bibitem{scheidegger2019constrained}
F.~Scheidegger, L.~Benini, C.~Bekas, and A.~C.~I. Malossi, ``Constrained deep
  neural network architecture search for {IoT} devices accounting for hardware
  calibration,'' in \emph{Advances in Neural Information Processing Systems},
  2019, pp. 6054--6064.

\bibitem{kumagai2019transfer}
A.~Kumagai, T.~Iwata, and Y.~Fujiwara, ``Transfer anomaly detection by
  inferring latent domain representations,'' in \emph{Advances in Neural
  Information Processing Systems}, 2019, pp. 2467--2477.

\bibitem{sanchez2014improving}
D.~S{\'a}nchez, J.~Domingo-Ferrer, and S.~Mart{\'\i}nez, ``Improving the
  utility of differential privacy via univariate microaggregation,'' in
  \emph{International Conference on Privacy in Statistical Databases}, 2014,
  pp. 130--142.

\bibitem{raspberrypi}
\BIBentryALTinterwordspacing
``Raspberry {P}i 4 tech specs,'' 2020, accessed on August 13, 2020. [Online].
  Available:
  \url{https://www.raspberrypi.org/products/raspberry-pi-4-model-b/specifications/}
\BIBentrySTDinterwordspacing

\bibitem{bonawitz2019towards}
K.~Bonawitz, H.~Eichner, W.~Grieskamp, D.~Huba, A.~Ingerman, V.~Ivanov,
  C.~Kiddon, J.~Kone{\v{c}}n{\`y}, S.~Mazzocchi, H.~B. McMahan \emph{et~al.},
  ``Towards federated learning at scale: System design,'' \emph{arXiv preprint
  arXiv:1902.01046}, 2019.

\bibitem{yin2019dreaming}
H.~Yin, P.~Molchanov, J.~M. Alvarez, Z.~Li, A.~Mallya, D.~Hoiem, N.~K. Jha, and
  J.~Kautz, ``Dreaming to distill: Data-free knowledge transfer via
  deepinversion,'' in \emph{Proceedings of the IEEE/CVF Conference on Computer
  Vision and Pattern Recognition}, 2020, pp. 8715--8724.

\end{thebibliography}

\end{document}